\documentclass[10pt, final, journal, letterpaper, twocolumn]{IEEEtran}

\usepackage{acronym}
\usepackage{amsfonts}
\usepackage[dvips]{graphicx}
\usepackage{times}
\usepackage{cite}
\usepackage{amsmath}
\usepackage{array}
\usepackage{amssymb}
\usepackage{stfloats}
\usepackage{slashbox}
\usepackage{graphicx}
\usepackage{footnote}
\usepackage{amsthm}
\usepackage{booktabs}
\usepackage{array}
\usepackage{algorithmic}
\usepackage{algorithm}
\usepackage{subeqnarray}
\usepackage{cases}
\usepackage{threeparttable}
\usepackage{color}
\usepackage{epstopdf}
\usepackage{multirow}
\usepackage{tabularx}
\usepackage{enumerate}
\usepackage{multicol}

\newtheorem{corollary}{Corollary}

\newtheorem{lemma}{Lemma}

\newtheorem{proposition}{Proposition}

\newtheorem{definition}{Definition}

\newcommand{\figref}[1]{{Fig.}~\ref{#1}}

\def\bb0{{\mathbb{0}}}

\def\bb{{\mathbf{b}}}

\def\br{{\mathbf{r}}}

\def\bx{{\mathbf{x}}}

\def\b0{{\mathbf{0}}}

\def\bH{{\mathbf{H}}}

\def\cH{\mathcal{H}}

\def\sf0{{\mathsf{0}}}

\def\rm0{{\mathrm{0}}}

\def\Nt{{N_\mathrm{t}}}
\def\Nr{{N_\mathrm{r}}}

\def\Pt{{P_{\mr{t}}}}
\def\Hb{{\mathcal{H}_{\mr{b}}}}

\newcommand{\mb}{\mathbf}
\newcommand{\mr}{\mathrm}
\def\j{\mathrm{j}}
\def\Re{\mathbf{Re}}
\def\Im{\mathbf{Im}}
\acrodef{CSI}[CSI]{channel state information}
\acrodef{CSIT}[CSIT]{channel state information at the transmitter}
\acrodef{CSIR}[CSIR]{channel state information at the receiver}
\acrodef{MIMO}[MIMO]{multiple-input multiple-output}
\acrodef{SISO}[SISO]{single-input single-output}
\acrodef{MISO}[MISO]{multiple-input single-output}
\acrodef{SIMO}[SIMO]{single-input multiple-output}
\acrodef{ADCs}[ADCs]{analog-to-digital convertors}
\acrodef{SNR}[SNR]{signal-to-noise ratio}
\acrodef{AWGN}[AWGN]{additive white Gaussian noise}
\acrodef{MRT}[MRT]{maximal ratio transmission}

\pdfoutput=1

\begin{document}
\title{Capacity Analysis of One-Bit Quantized MIMO Systems with Transmitter Channel State Information}
\author{\authorblockN{Jianhua~Mo,~\IEEEmembership{Student~Member,~IEEE}, and Robert W. Heath, Jr., ~\IEEEmembership{Fellow,~IEEE}}\\
\thanks{The authors are with Wireless Networking and Communications Group, The University of Texas at Austin, Austin, TX 78712, USA (Email: \{jhmo, rheath\}@utexas.edu.)}
\thanks{The material in this paper was presented in part at the 2014 Information Theory and Applications Workshop \cite{Mo_Jianhua_ITA14}.}
\thanks{This work is supported in part by the National Science Foundation under Grants No. 1218338 and 1319556. }}
\maketitle

\begin{abstract}
With bandwidths on the order of a gigahertz in emerging wireless systems, high-resolution analog-to-digital convertors (ADCs) become a power consumption bottleneck. One solution is to employ low resolution one-bit ADCs. In this paper, we analyze the flat fading multiple-input multiple-output (MIMO) channel with one-bit ADCs. Channel state information is assumed to be known at both the transmitter and receiver. For the multiple-input single-output channel, we derive the exact channel capacity.
For the single-input multiple-output and MIMO channel, the capacity at infinite signal-to-noise ratio (SNR) is found. We also derive upper bound at finite SNR, which is tight when the channel has full row rank.
In addition, we propose an efficient method to design the input symbols to approach the capacity achieving solution.
We incorporate millimeter wave channel characteristics and find the bounds on the infinite SNR capacity.
The results show how the number of paths and number of receive antennas impact the capacity.
\end{abstract}


\section{Introduction}
Bandwidth and antennas are growing in next generation wireless systems. A main reason is due to the use of millimeter wave (mmWave) carrier frequencies in personal area networks \cite{Baykas_COMM11}, local area networks \cite{Ghosh_JSAC14}, and likely even cellular networks \cite{Rappaport_Access13}. The channel bandwidths in mmWave systems are larger than those used in lower frequency UHF (ultra high frequency) systems. For example, in IEEE 802.11ad the bandwidth is $2.16$ GHz while in potential mmWave cellular applications, bandwidths of $500$ MHz or more are being considered \cite{Hong_Wonbin_COMM14}. Antenna arrays are important for mmWave systems. They provide array gain that helps mmWave systems achieve a favorable link margin. Due to the small carrier wavelength, a large number of co-located antennas can be deployed within a fixed antenna area in mmWave systems. For example, some IEEE 802.11ad chipsets use $32$ antennas \cite{Wilocity}, while mmWave celular applications envision hundreds of antennas at the base station and perhaps a dozen on the handset \cite{Hong_Wonbin_COMM14}. This motivates the study of MIMO systems with large numbers of antennas and higher channel bandwidths.

High bandwidth channels introduce new challenges in system design compared with design at lower frequencies \cite{Rangan_IEEE14}. One major issue is the power consumption associated with  the analog-to-digital conversion.
In conventional \ac{MIMO} receiver designs, the \ac{ADCs} are expected to have high resolution (e.g., more than $8$ bits) and act as transparent waveform preservers.
In channels with larger bandwidths, the corresponding sampling rate of the \ac{ADCs} scales up. Unfortunately, high-speed (e.g., more than  $1$ GSample/s), high-resolution ADCs are costly and power-hungry for portable devices\cite{Walden_JSAC99, Murmann_13, Le_Bin_SPM05}.
For example, in an ideal $b$-bit ADC with flash architecture, there are $2^b-1$ comparators and therefore the power consumption grows exponentially with the resolution \cite{Walden_JSAC99}. At present, commercially available \ac{ADCs} with high-speed and high-resolution consume power on the order of several Watts\cite{TI_datasheet}. Furthermore, because most communication systems exploit some form of \ac{MIMO} operation, multiple ADCs will be needed to quantize the received signals from multiple antennas separately if conventional digital baseband processing of all antenna outputs is assumed. Therefore, the total power assumption can be excessive, especially at the mobile station.

The most direct solutions to the power assumption bottleneck are to reduce the sampling rate and/or the quantization resolution of ADCs.
The sampling rate can be reduced by employing a special ADC structure called the time-interleaved ADC (TI-ADC) where a number of low-speed, high-resolution ADCs operate in parallel. The main drawback of the TI-ADC is the mismatch among the sub-ADCs in gain, timing and voltage offset which can cause error floors in receiver performance (though it is possible to compensate the mismatch at the price of higher complexity of the receiver \cite{Vitali_TCS09, Ponnuru_TCOM10}).
An alternative to high resolution ADCs is to live with ultra low resolution ADCs (1-3 bits), which reduces power consumption and cost.
The use of low resolution and especially one-bit ADCs radically changes both the theory and practice of communication. For example, the capacity maximizing transmit signals are discrete \cite{Witsenhausen_IT80, Singh_TCOM09}. This is in contrast with unquantized MIMO systems where the optimal input distribution is continuous. Although it is finite-dimensional, finding the optimal discrete input distribution is a challenging problem that depends on the \ac{CSIT} and \ac{CSIR}.
In \cite{Singh_TCOM09}, the capacity of the real-valued \ac{SISO} channel with \ac{CSIT} and \ac{CSIR} was considered. For one-bit quantization, binary antipodal signaling was found to be optimal. It was also shown that at low SNR, the use of low-resolution ADCs incurs a surprisingly small loss in spectral efficiency compared to unquantized observations.
In the block fading SISO channel without \ac{CSIT} and \ac{CSIR}, it was proven in \cite{Mezghani_ISIT08} that the capacity is achieved by on-off QPSK signaling where the on-off probability depends on the coherence time.
The results in \cite{Singh_TCOM09} and \cite{Mezghani_ISIT08}, however, do not readily extend to the MIMO channel.

The most related work to our contribution is \cite{Mezghani_ISIT07} where the one-bit quantized \ac{MIMO} channel with only \ac{CSIR} was considered. It was found that under the constraint that each antenna transmits signals independently and with equal power, independent QPSK signaling across different antennas is optimal in the low SNR regime.
In addition, there is a reduction of low SNR channel capacity by a factor $2/\pi$ ($-1.96$ dB) due to one bit quantization \footnote{This type of quantization loss is also well known to the error control coding community. At low SNR, there exists roughly a 2-dB difference between the performance of hard and soft decision decoding.
See, for example, \cite[Section 7.6]{Proakis_Book08} for details.}.
The results, unfortunately do not apply to medium or high SNR regimes.
Hence remains of interest to provide a complete characterization of the high SNR capacity in quantized MIMO systems.

In \cite{Murray_VTC06, Nossek_IWCMC06, Ivrlac_ISIT06}, the capacity of the quantized MIMO channel was derived under different assumptions. The input symbols, though, were only BPSK or QAM symbols since they assumed there is no \ac{CSI} at the transmitter with which to optimize the constellation. Therefore, what is called \emph{capacity} in \cite{Murray_VTC06, Nossek_IWCMC06, Ivrlac_ISIT06} is actually \emph{the mutual information} or \emph{an achievable rate} since the input distributions are not optimized. Though reasonable for practical implementation, without a carefully designed input distribution, the mutual information of quantized MIMO channel may achieve its maximum at a finite SNR (this phenomenon is called \emph{stochastic resonance}). With CSIT, the transmitter can implement beamforming to obtain array gain and can also design the transmitted constellation to achieve higher achievable rates.
In this paper, we consider the quantized channel with both perfect \ac{CSIT} and \ac{CSIR}. Algorithms for CSI estimation in the one-bit framework have been considered in related work \cite{Dabeer_ICC10, Zeitler_TSP12, Lok_ISIT98, Ivrlac_WSA07, Mezghani_WSA10, Mezghani_WSA12, Jacques_IT13, Wen_Chao-Kai_arxiv15} and our work \cite{Mo_Jianhua_Asilomar14}. Incorporating estimation error and studying CSI feedback are left to future work.

For low-resolution ADCs with more than one-bit resolution, i.e., 2-3 bits, besides the difficulty of finding the optimal input distribution, another major challenge is selecting the appropriate quantizer thresholds.
Although the well-known uniform quantizer and nonuniform Lloyd-Max quantizer are commonly used for receiver design and rate analysis in related work (e.g., \cite{Murray_VTC06, Bai_Qing_ETT14, Mezghani_MELECON12, Mezghani_WSA12, Wang_Shengchu_TWC15, Orhan_ITA15, Wen_Chao-Kai_arxiv15}),
they are not necessarily optimal for maximizing the channel capacity. For example, assuming Lloyd-Max quantizer is used and the quantization error is treated as Gaussian noise, a lower bound of the channel capacity is derived in \cite{Mezghani_WSA07, Bai_Qing_ETT14}. The lower bound is tight only at low SNR.
In addition, the two challenges are coupled, which makes this problem even difficult.
Namely, the optimal input distribution depends on the thresholds of ADCs; the thresholds of ADCs will also affect the choice of input distribution. Furthermore, the input distribution and the thresholds may change greatly with SNR (see Fig. 5 in \cite{Singh_TCOM09} as an example). In the narrowband \ac{SISO} channel, a numerical approach is to iteratively optimize the input distribution and the ADC thresholds \cite{Singh_TCOM09}. For the SISO frequency-selective channel, vector quantization may be adopted to better exploit the correlation in the received sequence \cite{Zeitler_TCOM12}.
Related work \cite{Lu_Minwei_ISCAS10, Narasimha_TSP12} on the SISO frequency-selective channel also optimized the thresholds numerically, but based on a bit-error-rate criterion.


In this paper, we focus on the MIMO channel with one-bit quantization. The main advantage of this architecture is that the one-bit ADC is just a simple comparator and can be implemented with very low power consumption \cite{Sundstrom_TCASI09}. The architecture also simplifies the overall complexity of the circuit, for example automatic gain control (AGC) may not be required \cite{Singh_ICUWB09}.
Compared to our initial work \cite{Mo_Jianhua_ITA14}, a tighter bound on the MIMO channel capacity at infinite SNR was provided in this paper. In addition, new bounds on capacity at finite SNR are included in this paper.

The main contributions of this paper are summarized as follows.
\begin{itemize}
  \item We find the capacity of the \ac{MISO} channel with one-bit quantization in the whole SNR regime.
  \item We derive the infinite SNR capacity of the \ac{SIMO} channel with one-bit quantization. We also find a closed-form expression when the receiver has a large number of antennas.
  \item We provide bounds on the infinite SNR capacity of the MIMO channel with one-bit quantization. The decoding process at the receiver is similar to
  finding the transmitted symbols satisfying a series of linear inequalities.
  Based on this observation, we develop accurate bounds on the infinite SNR capacity by relating it to a problem in classical combinatorial geometry.
  A computationally efficient method based on convex optimization is proposed to design the input alphabet such that the infinite SNR capacity is approached.
  \item We provide a new upper bound for the \ac{MIMO} channel with one-bit quantization at finite SNR. The bound is tight when the channel is row full rank. A simple lower bound by using channel inversion strategy is derived. We also prove that the lower bound obtained by treating the quantization noise as Gaussian noise is loose at high SNR.
  \item We find the infinite SNR capacity of a sparse mmWave channel. We show that the capacity is mainly limited by the number of paths in the mmWave propagation environment. In a special case when there is only one single path, we propose a capacity-achieving transmission strategy.
\end{itemize}

The paper is organized as follows. In Section II, we describe a MIMO system with one-bit quantization. In Section III, we present the SISO and MISO capacities of the one-bit quantized MIMO channel. In Section IV and V, we analyze the capacity of MIMO channel at infinite and finite SNR, respectively. Numerical methods to optimize the distribution of input symbols are shown in Section VI. We then consider the quantized mmWave MIMO channel in Section VII. Simulation results are shown in Section VIII, followed by the conclusions in Section IX.

\emph{Notation} : $a$ is a scalar, $\mb{a}$ is a vector and $\mb{A}$ is a matrix. $\angle x$ represents the phase of a complex number $x$.  $\mb{x}_{i:j}$ is the vector consisting of \{$x_k$, $i \leq k \leq j$\}. $\mr{tr}(\mb{A})$, $\mb{A}^T$ and $\mb{A}^*$ represent the trace, transpose and conjugate transpose of a matrix $\mb{A}$, respectively.
$\mb{A \odot \mb{}B}$ stands for the Hadamard product of $\mb{A}$ and $\mb{B}$. $\mr{diag}(\mb{a})$ represents a square diagonal matrix with the elements of vector $\mb{a}$ on the diagonal. $\Re(x)$ and $\Im(x)$ stand for the real and imaginary part of $x$, respectively. $\mr{Pr}[\cdot]$ denotes the probability.

\section{System Model}

\begin{figure}[t]
\begin{centering}
\includegraphics[scale=.45]{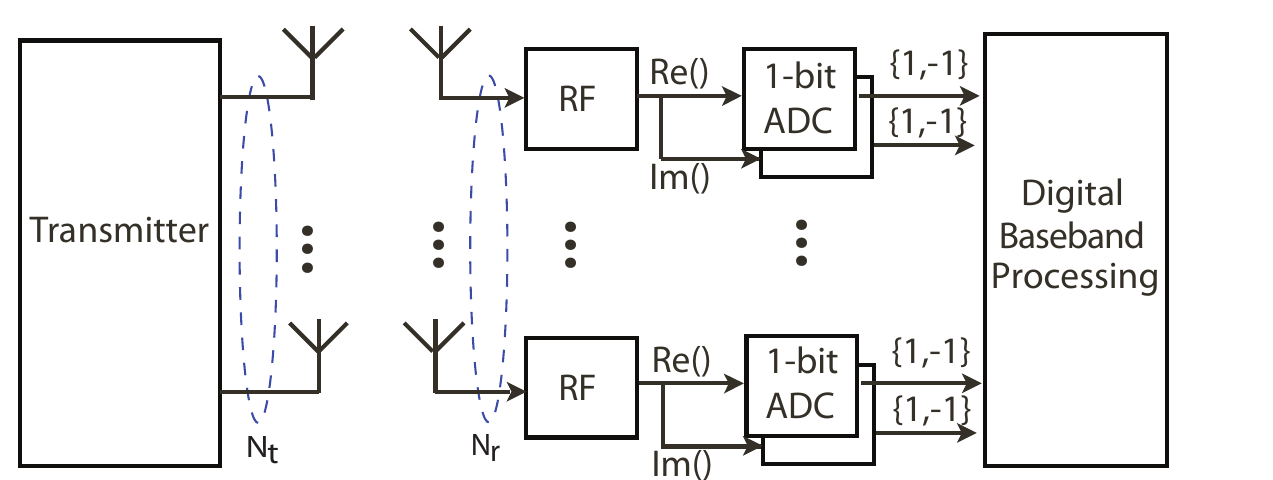}
\vspace{-0.1cm}
\centering
 \caption{A $\Nr \times \Nt$ MIMO system with one-bit quantization at the receiver.  For each receiver antenna, there are two one-bit ADCs. Note that there is no limitation on the structure of the transmitter.}\label{fig:System_model}
\end{centering}
\vspace{-0.3cm}
\end{figure}

Consider a MIMO system with one-bit quantization, as shown in Fig. \ref{fig:System_model}.
There are $\Nt$ antennas at the transmitter and $\Nr$ antennas at the receiver.
Assuming perfect synchronization and a narrowband channel, the baseband received signal in this $\Nr \times \Nt$ MIMO system is,
\begin{equation}
  \mb{y} = \mb{H}\mb{x}+\mb{n}
\end{equation}
where $\mb{H} \in \mathbb{C}^{\Nr\times \Nt}$ is the channel matrix, $\mb{x}\in \mathbb{C}^{\Nt \times 1}$ is the signal sent by the transmitter, $\mb{y}\in \mathbb{C}^{\Nr \times 1}$ is the received signal before quantization, and $\mb{n} \sim \mathcal{CN}(0, \mb{I})$ is the circularly symmetric complex Gaussian noise.

In our system, there are  a total of $2\Nr$ one-bit resolution quantizers that separately quantize the real and imaginary part of each received signal.
The output after the one-bit quantization is
\begin{equation}
  \mb{r} = \mr{sgn}\left(\mb{y}\right) = \mr{sgn}\left(\mb{H}\mb{x}+\mb{n}\right),
\end{equation}
where $\mr{sgn}()$ is the signum function applied componentwise and separately to the real and imaginary parts. Therefore, the quantization output at the $i$th antenna $r_i \in \{1+\j, 1-\j, -1+\j, -1-\j\}$ for $1\leq i \leq \Nr$.

In this paper, we assume that there is both \ac{CSIT} and \ac{CSIR}. Consequently, the channel capacity with one-bit quantization is
\begin{eqnarray} \label{eq:def_MIMO_Capacity}
  C =  \max_{p(\mb{x}): \mr{tr}\left(\mathbb{E} \left(\mb{x} \mb{x}^* \right)\right)\leq \Pt}I(\mb{x};\mb{r}| \mb{H}),
\end{eqnarray}
where
\begin{eqnarray} \label{eq:def_MI}
  I(\mb{x}; \mb{r}| \mb{H}) = \int_{\mb{x}} \sum_{\mb{r}} \mr{Pr}(\mb{x}) \mr{Pr}(\mb{r}|\mb{x}) \log_2 \frac{\mr{Pr}(\mb{r}|\mb{x})}{\mr{Pr}(\mb{r})} \mr{d} \mb{x},
\end{eqnarray}
and $\Pt$ is the average power constraint at the transmitter.


\section{SISO and MISO Channel Capacities with One-bit Quantization}
The mutual information in \eqref{eq:def_MI} is in the form of multiple integrals and sums. Obtaining a closed form expression for the optimization problem is a challenge.
For the simple \ac{SISO} and \ac{MISO} cases where there is only one receive antenna, we find the capacity and the capacity-achieving strategy. We start with the \ac{SISO} case then move on to the \ac{MISO} case.

\subsection{SISO Channel with One-Bit Quantization}
First, we deal with the very special case when $\Nt=\Nr=1$. The channel coefficient now is a scalar denoted by $h$.
\begin{lemma} \label{lemma:SISO}
The capacity of the SISO channel with one-bit quantization is
\begin{equation}
  C_{\mr{1bit, SISO}} = 2\left(1 - \Hb \left(  Q\left(\left|h\right| \sqrt{\Pt}\right) \right)\right),
\end{equation}
where $\Hb(p) = -p \log_2 p -(1-p) \log_2 (1-p)$ is the binary entropy function, $Q(\cdot)$ is the tail probability of the standard normal distribution, and
the capacity is achieved by rotated QPSK signaling with uniform probabilities, i.e.,
\begin{equation}
  \mr{Pr}\left[x = \sqrt{\Pt} e^{\j\left( \frac{k \pi}{2} + \frac{\pi}{4} - \angle h \right)}\right] = \frac{1}{4}, \; \mr{for} \; k=0, 1, 2 \;\mr{and}\; 3.
\end{equation}
\end{lemma}
\begin{IEEEproof}
     Without loss of optimality, we can assume that the transmitted signal is $x = e^{-\j \angle h} \hat{x}$. The outputs of the one-bit quantizer will be $\Re(r) = \text{sgn} \left(|h|\Re(\hat{x}) + \Re(n) \right)$ and $\Im(r) = \text{sgn} \left(|h|\Im(\hat{x}) + \Im(n) \right)$. Hence the channel is decoupled into two real-valued channels with the same channel gain. For the real-valued channel, it was proven in \cite[Theorem 2]{Singh_TCOM09} that binary antipodal signaling is optimal and the capacity is $1 - \Hb \left(  Q\left(\left|h\right| \sqrt{\Pt}\right) \right)$. In addition, to maximize the sum of the capacities of these two real-valued channels, the total should be split equally since $\Hb\left( Q\left( \sqrt{x}\right)\right)$ is a convex function \cite{Dabeer_SPAWC06}.
     Hence, the optimal input for the SISO channel is rotated QPSK signaling and the channel capacity is $C_{\mr{1bit, SISO}} = 2\left(1 - \Hb \left(  Q\left(\left|h\right| \sqrt{\Pt}\right) \right)\right)$.
\end{IEEEproof}

Note that the capacity of the one-bit quantized SISO channel is similar to that of the binary symmetric channel with crossover probability $Q\left(\left|h\right| \sqrt{\Pt}\right)$, which is $1 - \Hb \left(  Q\left(\left|h\right| \sqrt{\Pt}\right) \right)$ \cite[Section 7.1.4]{Cover_Book12}.


\subsection{MISO Channel with One-Bit Quantization}
In this subsection, we consider the MISO channel with one bit quantization. The received signal is
    \begin{eqnarray}
      r = \mr{sgn}\left( y \right) = \mr{sgn} \left(\mb{h}^* \mb{x} + n \right)
    \end{eqnarray}
where $\mb{h} \in \mathbb{C}^{\Nt \times 1}$ is the channel vector.
If ADCs with infinite resolution are used at the receiver (i.e., there is no quantization noise), the capacity-achieving transmission strategy is to use \ac{MRT} beamforming and Gaussian signaling. In a system with one-bit ADCs, Gaussian signaling, it turns out, is not optimal.

\begin{proposition}
The capacity of the MISO channel with one-bit quantization is
\begin{equation}
  C_{\mr{1bit, MISO}} = 2\left(1 - \Hb\left(  Q\left(||\mb{h}|| \sqrt{\Pt}\right) \right)\right),
\end{equation}
and the capacity is obtained by \ac{MRT} beamforming and QPSK signaling, i.e.,
\begin{equation}
  \mr{Pr}\left[\mb{x} =  \sqrt{\Pt} \frac{\mb{h}}{||\mb{h}||} e^{\j \left( \frac{k \pi}{2} + \frac{\pi}{4} \right)}\right] = \frac{1}{4}, \; \mr{for} \; k=0, 1, 2 \; \mr{and} \; 3.
\end{equation}
\end{proposition}

\begin{IEEEproof}
Assume that the transmitted symbol is $\mb{x} = \mb{U} \mb{s}$ where $\mb{U} \in \mathbb{C}^{\Nt \times \Nt}$ is a unitary matrix and $\mb{s} = [s_1, s_2, \cdots, s_{\Nt}]^T \in \mathbb{C}^{\Nt \times 1}$ is the information-bearing signal. Since $\mb{s} \rightarrow \mb{U} \mb{s}$ is one-to-one mapping, $I(\mb{s}; r)=I(\mb{Us};r) = I(\mb{x};r)$. Therefore, this assumption does not change the capacity.
Assuming the unitary matrix $\mb{U}=\left[\frac{\mb{h}}{||\mb{h}||},  \overline{\mb{U}}_{\Nt \times (\Nt-1)} \right]$, then $r = \mr{sgn}(|| \mb{h} || s_1 + n)$ where $s_1$ is the first element in $\mb{s}$. Therefore, the MISO channel is transformed to an equivalent SISO channel with channel gain $||\mb{h}||$. To maximize the mutual information, we set $s_2 =s_3= \cdots = s_{\Nt}=0$ and $\mathbb{E}[s_1s_1^*]=\Pt$. Therefore, $\mb{x} = \frac{\mb{h}}{||\mb{h}||}s_1$.
Then Proposition 1 follows by using Lemma 1.
\end{IEEEproof}


Similar to the SISO case, $C_{\mr{1bit, MISO}}$ converges to the upper bound 2 bps/Hz in the high SNR regime.
In the low SNR regime,
\begin{eqnarray}
  & & C_{\mr{1bit, MISO}} \nonumber \\
  &\stackrel{(a)} = &  2 \left(1 - \Hb \left(\frac{1}{2} - \frac{1}{\sqrt{2 \pi}} ||\mb{h}|| \sqrt{\Pt} + o(||\mb{h}||^2 \Pt)\right)\right) \nonumber \\
  &\stackrel{(b)} = & 2 \left( 1 - \left( 1 - \frac{2}{\ln 2}  \left( \frac{1}{\sqrt{2 \pi}} ||\mb{h}|| \sqrt{\Pt} \right)^2  \right. \right. \nonumber \\
  & & \left. \left. + \frac{4}{3 \ln 2} \left( \frac{1}{\sqrt{2 \pi}} ||\mb{h}|| \sqrt{\Pt} \right)^4 + o\left( ||\mb{h}||^4 \Pt^2 \right) \right) \right) \nonumber \\
  &=& \frac{2}{\pi} \frac{||\mb{h}||^2 \Pt}{\ln 2} - \frac{2}{3 \pi^2} \frac{||\mb{h}||^4 \Pt^2} {\ln 2 } + o\left( ||\mb{h}||^4 \Pt^2 \right) \label{eq:MISO_approx}
\end{eqnarray}
where (a) and (b) follow from $Q(t) = \frac{1}{2} - \frac{1}{\sqrt{2\pi}} t + o(t^2)$ and $\Hb(\frac{1}{2}+t) = 1 - \frac{2}{\ln 2} t^2 + \frac{4}{3 \ln 2} t^4 + o(t^4)$, respectively. Note that in the low SNR regime, the capacity of MISO channel without quantization is $C_{\mr{MISO}} = \log_2 \left(1+ ||\mb{h}||^2 \Pt \right)= \frac{||\mb{h}||^2 \Pt}{\ln 2} + o(\Pt)$. Therefore, in the MISO channel with CSIT, the one-bit quantization results in $10 \log_{10} \frac{\pi}{2}=1.96$ dB power loss. A similar result was reported in \cite{Mezghani_ISIT07} but under the assumption of only CSIR.

When there is only \ac{CSIR}, as shown in \cite[Theorem 2]{Mezghani_ISIT07}, the achievable rate with independent QPSK signaling on each transmitter antenna is
\begin{eqnarray} \label{eq:MISO_qpsk_approx}
  R_{\mr{1bit, MISO}}^{\mr{QPSK}} = \frac{2}{\pi} \frac{||\mb{h}||^2 \Pt}{\Nt \ln 2} + o\left( ||\mb{h}||^2 \Pt \right).
\end{eqnarray}
Comparing the first terms in \eqref{eq:MISO_approx} and \eqref{eq:MISO_qpsk_approx}, we can see that independent QPSK signaling results in $1/{\Nt}$ power loss compared to the optimal strategy.
The reason is that the optimal strategy has the array gain from beamforming.

\section{SIMO and MIMO Channel Capacities at Infinite SNR with One-Bit Quantization}\label{sec:infinite_SNR}
For SIMO and MIMO channel with one-bit quantization, the exact capacity is unknown. In this section, we consider the special case when the SNR is infinite, i.e., there is no additive white Gaussian noise. In the next section, we will provide bounds on the capacity at finite SNR.

\subsection{SIMO Channel with One-Bit Quantization}
First we consider the SIMO channel. With $\Nr$ antennas at the receiver, there are at most $2^{2\Nr}$ possible quantization outputs. Therefore, $2 \Nr$ bps/Hz is a simple upper bound on the channel capacity. This upper bound, unfortunately, cannot be approached when $\Nr$ is larger than one. We provide a precise characterization in the following proposition.

\begin{proposition} \label{prop:SIMO}
The capacity of the SIMO channel with one-bit quantization at infinite SNR, denoted as $\overline{C}_{\mr{1bit, SIMO}}$, satisfies
\begin{equation} \label{eq:Rate_SIMO}
  \log_2 (4 \Nr) \leq \overline{C}_{\mr{1bit, SIMO}} \leq \log_2 \left(4 \Nr + 1\right).
\end{equation}
\end{proposition}
\begin{IEEEproof}
Denote the SIMO channel as $\mb{h}=[h_1 , h_2 , \cdots , h_{\Nr}]^T$. When the phase of the transmitted symbol $x$ is around $\angle x = k \pi/2 - \angle{h_i} \; (k=0,1,2,3; \; i=1,2,\cdots,\Nr)$, one element of the one-bit quantization output will change. There are at most $4 \Nr$ such phases, denoted as $\mr{\Phi} = \{\phi_i, 1\leq i \leq 4\Nr\}$. We usually assume that the channel coefficients are generated from continuous distribution. Then with probability one, $\angle h_m \neq \angle h_n + k \pi/2, k\in \{0, 1, 2, 3\}$ for $m \neq n$. Therefore, we assume there are $4 \Nr$ different phases.

In \figref{fig:SIMO_signaling}, we show such an example when $\Nr=3$. The 6 axes represent the quantization thresholds of the 6 one-bit ADCs seen at the transmitter. The axes are offset due to the rotations induced by the channel coefficients. It is shown that the entire plane is divided into 12 regions.

The possible transmit symbols on the complex plane can be divided into three categories:
  \begin{enumerate}
    \item The symbol zero, i.e., $x=0$;
    \item The symbols with phases in $\mr{\Phi}$ (i.e., the input symbols on the axe);
    \item The symbols  with phases not in $\mr{\Phi}$ (for instance, the input symbols shown in \figref{fig:SIMO_signaling} except the symbol zero).
  \end{enumerate}

\begin{figure}[t]
\begin{centering}
\includegraphics[scale=.38]{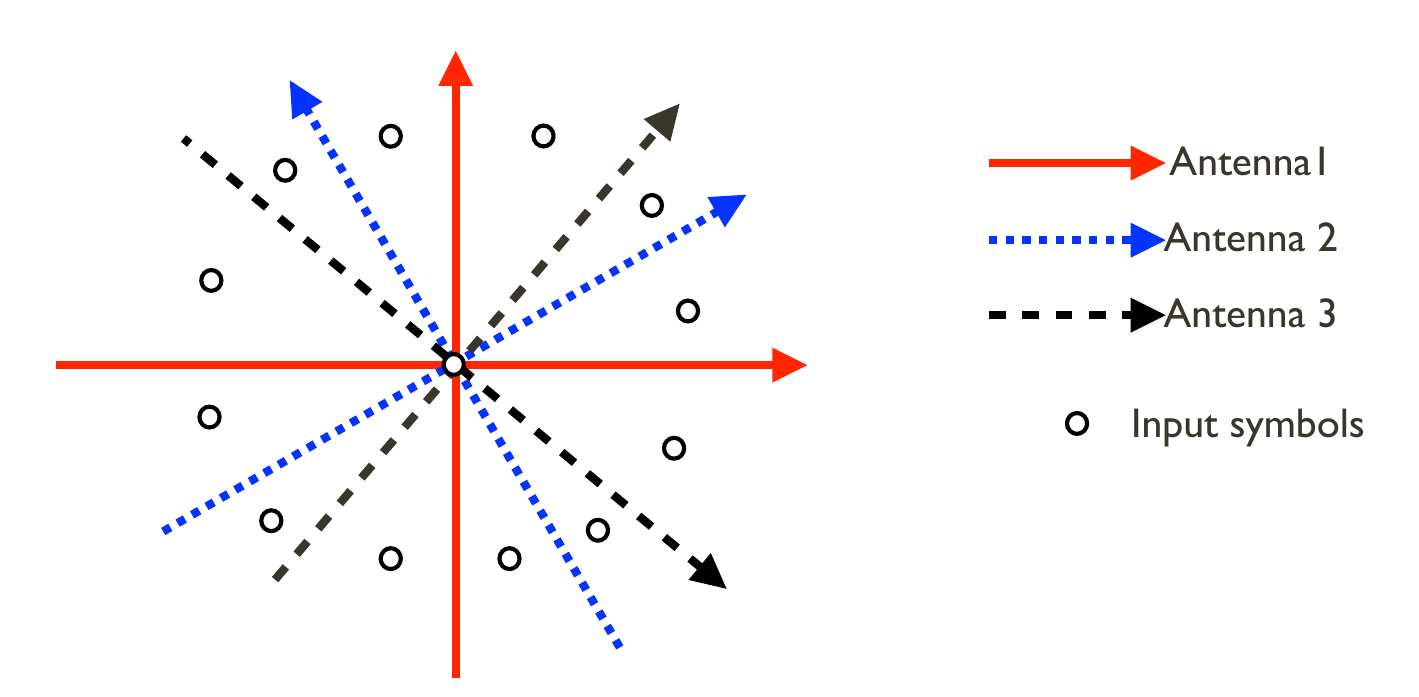}
\vspace{-0.1cm}
\centering
 \caption{The transmitted symbols of a SIMO channel with 3 receive antennas in the high SNR regime. Here, $\angle h_1=0$ and $\angle h_2 \neq \angle h_3 \neq 0$.
 The optimal constellation contains 12 nonzero symbols and the symbol zero.}\label{fig:SIMO_signaling}
\end{centering}
\vspace{-0.5cm}
\end{figure}

  For the zero symbol, the transition probability $\mr{Pr}\left[ \mb{r}|x=0 \right]$ is $2^{-2\Nr}$ for each possible $\mb{r}$.
  For the symbols with phases not in $\Phi$, $\mr{Pr}\left[\mb{r}|x \right] = \mathbf{1}_{\left\{\mb{r}=\mb{Q}(\mb{h}x) \right\}}$ where $\mathbf{1}_{\left\{\cdot \right\}}$ is the indicator function.
   At last consider the symbols with phases in $\Phi$. If $\angle x = -\angle h_1$, then $\mr{Pr}\left[r_1=1+\j \;|x \right] =\mr{Pr} \left[r_1=1-\j \;|x \right]=1/2$. The other conditional probabilities can be derived similarly.
  Therefore, the transition probability matrix, which has the dimension as $(8 \Nr+1) \times 2^{2\Nr}$, is
 \begin{eqnarray}
 \mr{Pr} \left[ \mb{r}|x \right] = \left[
             \begin{array}{ccc}
               {2^{-2\Nr}} \mb{1}_{1 \times 4 \Nr}  &  & 2^{-2\Nr} \mb{1}_{1 \times (2^{2\Nr}-4\Nr)} \\
               \mb{T}_{4\Nr\times 4\Nr}             &  & \mb{0}_{4\Nr \times (2^{2\Nr}-4\Nr)} \\
               \mb{I}_{4\Nr\times 4\Nr} &  & \mb{0}_{4\Nr \times (2^{2\Nr}-4\Nr)}
             \end{array}
           \right]
 \end{eqnarray}
where all the entries of the first row are $2^{-2 \Nr} $ and
$\mb{T}$ is a $4\Nr \times 4\Nr$ circulant matrix with the first row being $[1/2, 1/2, 0, \cdots, 0]$.

Assume that these three kinds of symbols are transmitted with probabilities $p_{\mr{0}}$, $p_{\mr{1}}$ and $1-p_{\mr{0}}-p_{\mr{1}}$, respectively. The resulting mutual information is,
  \begin{eqnarray}\label{eq:SIMO_mutual_information}
    f(p_{\mr{0}}, p_{\mr{1}}) & \triangleq & \left( { - 1 + {p_{\mr{0}}} - {p_{\mr{1}}} - {p_{\mr{0}}}\frac{{4{\Nr}}}{{{4^{{\Nr}}}}}} \right) \nonumber \\
 & & \times \log_2 \left( {\frac{{1 - {p_{\mr{0}}} - {p_{\mr{1}}}}}{{4{\Nr}}} + \frac{{{p_{\mr{0}}}}}{{{4^{{\Nr}}}}} + {p_{\mr{1}}}} \right) - 2{p_{\mr{1}}} \nonumber \\
 & &  - \frac{{8\Nr^2}}{{{4^{{\Nr}}}}}{p_{\mr{0}}} - \frac{{{4^{{\Nr}}} - 4{\Nr}}}{{{4^{{\Nr}}}}}{p_{\mr{0}}}\log_2 {p_{\mr{0}}}.
 \end{eqnarray}
The channel capacity can be computed by searching the optimal $p_{\mr{0}}$ and $p_{\mr{1}}$, denoted as $p_{\mr{0}}^*$ and $p_{\mr{1}}^*$, which maximizes the mutual information $f(p_{\mr{0}}, p_{\mr{1}})$. It turns out that ${\partial f(p_{\mr{0}}, p_{\mr{1}})}/{\partial p_{\mr{1}}}<0$ and thus $p_{\mr{1}}^*=0$. Therefore, there are at most $4 \Nr+ 1$ possible input symbols in the capacity-achieving distribution; an upper bound on the capacity is $\log_2 (4\Nr +1)$.

In \figref{fig:SIMO_signaling}, we show such an example when $\Nr=3$.
The optimal constellation contains 12 nonzero symbols falling in the 12 regions and the symbol zero.

The lower bound of $\log_2 (4 \Nr)$ is achieved by setting $p_{\mr{0}}=0$ and $p_{\mr{1}}=0$, i.e., $f(0, 0) = \log_2 (4 \Nr)$.
In other words, the lower bound is achieved by transmitting the distinguishable $4\Nr$ symbols with equal probability.
\end{IEEEproof}

\begin{corollary}
  When $\Nr$ is large, the capacity of the SIMO channel with one-bit quantization at infinite SNR is
  \begin{eqnarray}
    \overline{C}_{\mr{1bit, SIMO}} \approx \log_2(4 \Nr+1).
  \end{eqnarray}
\end{corollary}

\begin{IEEEproof}
In the proof of Proposition \ref{prop:SIMO}, we know that the optimal $p_{\mr{1}}$ is zero.
From \eqref{eq:SIMO_mutual_information}, the mutual information when $p_{\mr{1}}=0$ is,
  \begin{eqnarray}
   f(p_{\mr{0}}, 0)  & =& \left( { - 1 + {p_{\mr{0}}}  - {p_{\mr{0}}}\frac{{4{\Nr}}}{{{4^{{\Nr}}}}}} \right)\log_2 \left( {\frac{{1 - {p_{\mr{0}}}  }}{{4{\Nr}}} + \frac{{{p_{\mr{0}}}}}{{{4^{{\Nr}}}}} } \right)  \nonumber \\
   & & - \frac{{8\Nr^2}}{{{4^{{\Nr}}}}}{p_{\mr{0}}} - \frac{{{4^{{\Nr}}} - 4{\Nr}}}{{{4^{{\Nr}}}}}{p_{\mr{0}}}\log_2 {p_{\mr{0}}}.
\end{eqnarray}
When $\Nr$ is large, $\frac{\Nr}{4^{\Nr}} \rightarrow 0$ and $\frac{\Nr^2}{4^{\Nr}} \rightarrow 0$. Therefore,
\begin{eqnarray}
  f(p_{\mr{0}}, 0) & \approx & -(1-p_{\mr{0}}) \log_2 \frac{1-p_{\mr{0}}}{4 \Nr} - p_{\mr{0}} \log_2 p_{\mr{0}}.
\end{eqnarray}
It turns out that $p_{\mr{0}}^* = \frac{1}{4\Nr+1}$ and $f(p_{\mr{0}}^*, 0)\approx \log_2(4\Nr+1)$.
\end{IEEEproof}

\begin{figure}[t]
\begin{centering}
\includegraphics[scale=.6]{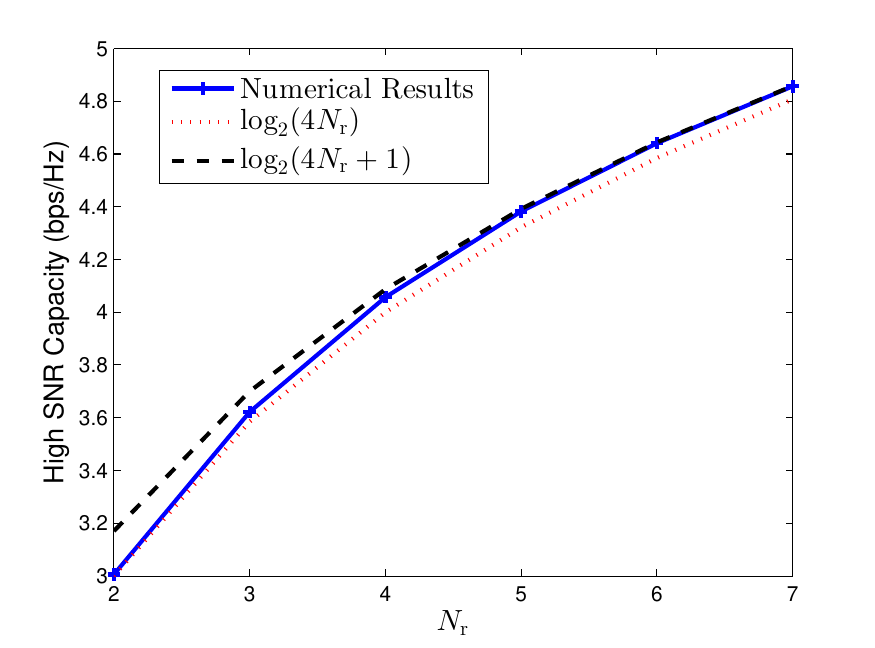}
\vspace{-0.1cm}
\centering
 \caption{The high SNR capacity and its lower and upper bounds. The high SNR capacity is very close to $\log_2(4 \Nr+1)$ when $\Nr \geq 6$. }\label{fig:SIMO_capacity_UB_LB}
\end{centering}
\vspace{-0.5cm}
\end{figure}

In Fig. \ref{fig:SIMO_capacity_UB_LB}, we plot the infinite SNR capacity obtained by numerically maximizing the mutual information function $f(p_{\mr{0}}, p_{\mr{1}})$.  The lower bound $\log_2 (4\Nr)$ and upper bound $\log_2 (4\Nr+1)$ are also plotted. It is shown that the high SNR capacity converges to $\log_2 (4\Nr+1)$ when $\Nr\geq 6$.

In communication systems, the zero symbol is often not included as part of the constellation due to the peak-to-average power ratio (PAPR) issue.
Therefore, in the high SNR regime, only $4 \Nr$ distinguishable symbols are employed as the channel inputs and there will be $4 \Nr$ different quantization outputs corresponding to each input symbol. The resulting achievable rate will converge to $\log_2 (4\Nr)$ as transmission power increases.

\subsection{MIMO Channel Capacity with One-Bit Quantization} \label{sec:MIMO_capacity}
In the infinite SNR regime, the decoding process at the receiver is as follows:
\begin{subequations}
\begin{eqnarray}
  \textbf{Find} & &\mb{x}  \\
   \text{s.t.} &  & \mr{sgn} (\mb{H} \mb{x}) = \mb{r}, \label{eq:MIMO_decoding}\\
   & & x \in \mathcal{X},
\end{eqnarray}
\end{subequations}
where $\mathcal{X}$ is the set containing all the input symbols and the equation in \eqref{eq:MIMO_decoding} is applied componentwise and separately to the real and imaginary parts.
Therefore, the decoding process is similar to finding the input symbols satisfying a system of linear inequalities,
\begin{eqnarray}
  \mb{r} \odot (\mb{H} \mb{x})>0. \label{eq:MIMO_decoding_inequalities}
\end{eqnarray}
where the Hadamard product and inequality are applied componentwise and separately to the real and imaginary parts.
If there is only one input symbol satisfying the system of linear inequalities, it can be correctly decoded at infinite SNR.
It turns out that the infinite SNR capacity of the quantized MIMO channel is closely related to a problem in classic combinatorial geometry.
We first give a related lemma and its dual statement; the proof is available in several references, e.g., \cite{Winder_SWCT61, Wendel_Math62, Cover_EC65, Flatto_AMS1970}.
\begin{lemma} \label{lemma:orthant}
     $N$ hyperplanes in general position passing through the origin of a $d$ dimensional space divide the space into
     $2 \sum_{k=0}^{d-1} \binom{N-1}{k}$ regions.

     \textbf{Dual:} A $d$ dimensional subspace in general position in $N$ dimensional space intersects $2 \sum_{k=0}^{d-1} \binom{N-1}{k}$ orthants.
\end{lemma}

We first give two examples of the lemma. First, a line passing through the origin divides any dimensional space into $2$ regions. Second, $N$ distinct lines passing through the origin divide the 2-D plane into $2 \binom{N-1}{0} + 2 \binom{N-1}{1}= 2 N$ regions.

\begin{definition}
A set of $N$ vectors is \emph{in general position} in $d$ dimensional space if and only if every subset of $d$ or fewer vectors is linearly independent.
\end{definition}
If we put the $N$ vectors, denoted as $\mb{w}_1, \mb{w}_2, \cdots, \mb{w}_{N}$, into a matrix $\mb{W}_{N \times d} = \left(\mb{w}_1, \mb{w}_2, \hdots, \mb{w}_N \right)^T$. These $N$ vectors are \emph{in general position} in $d$-dimensional space if and only if every $d \times d$ submatrix has a nonzero determinant \cite{Cover_EC65}. Note that general position is a strengthened rank condition\footnote{For example, the matrix $\left(
           \begin{array}{ccc}
             1 & 1  \\
             1 & 2  \\
             1 & 1
           \end{array}
         \right)$
 has full column rank but does not satisfy the condition of general position.}.

\begin{definition}
  A matrix $\mb{A}$ satisfies the condition of general position if and only if the set of row vectors is in general position.
\end{definition}

Based on Lemma \ref{lemma:orthant}, we provide our results on the MIMO channel capacity with one-bit quantization.

\begin{proposition} \label{prop:MIMO}
  If the channel matrix $\mb{H}$ satisfies the condition of general position, then the infinite SNR capacity satisfies
  \begin{equation} \label{eq:Rate_MIMO}
    \log_2 \left( K\left(\Nr, \Nt \right) \right)
    \leq \overline{C}_{\mr{1bit, MIMO}}
    \leq \log_2 \left( K\left( \Nr, \Nt \right) +1 \right)
  \end{equation}
   where
   \begin{eqnarray} \label{eq:def_K}
     K(\Nr, \Nt) &\triangleq & 2 \sum_{k=0}^{2\Nt-1} \binom{2\Nr-1}{k}
   \end{eqnarray}
   when $\Nt < \Nr$ and
   \begin{eqnarray}
     \overline{C}_{\mr{1bit, MIMO}} =  2\Nr
   \end{eqnarray}
   when $\Nt \geq \Nr$.
\end{proposition}

\begin{IEEEproof}
Assuming there is no noise, the equivalent real-valued channel is
\begin{eqnarray} \label{eq:notation_real}
  \widehat{\mb{r}} = \mr{sgn}(\widehat{\mb{H}} \widehat{\mb{x}}),
\end{eqnarray}
where $\widehat{\bf{x}}=\left[\Re(\mb{x})^T, \Im(\mb{x})^T\right]^T$, $\widehat{\mb{r}}=[\Re(\mb{r})^T , \Im(\mb{r})^T]^T$ and
\begin{eqnarray}
  \widehat{\mb{H}}_{2\Nr \times 2 \Nt} = \left[
             \begin{array}{cc}
               \Re(\mb{H})  & -\Im(\mb{H})  \\
               \Im(\mb{H}) & \Re(\mb{H}) \\
             \end{array}
           \right].
\end{eqnarray}

Each row of the channel matrix $\widehat{\mb{H}}$ defines a hyperplane passing through the origin of a $2\Nt$ dimensional space. These $ 2\Nr $ hyperplanes divide the $2\Nt$ dimensional space into $2 \sum_{k=0}^{2\Nt-1} \binom{2\Nr-1}{k}$ regions. For the dual statement, the subspace spanned by the $2 \Nt $ columns of $\widehat{\mb{H}}$ intersects $2 \sum_{k=0}^{2\Nt-1} \binom{2\Nr-1}{k}$ orthants in the $2\Nr$ dimensional space. In other words, there are $2 \sum_{k=0}^{2\Nt-1} \binom{2\Nr-1}{k}$ possible different $\mb{r}$'s by varying the transmitted symbol $\mb{x}$.

To achieve the channel capacity at infinite SNR, the transmitter has to send the zero symbol and $2 \sum_{k=0}^{2\Nt-1} \binom{2\Nr-1}{k}$ symbols from each region. The proof is similar to the proof of Proposition \ref{prop:SIMO}. Note that when $\Nt \geq \Nr$, the transmitter can send $2^{2 \Nr}$ symbols corresponding to all $2^{2 \Nr}$ possible quantization outputs with equal probabilities ${2^{-2 \Nr}}$. Hence, the zero symbol is not sent when $\Nt \geq \Nr$. Therefore, the proposition follows.
\end{IEEEproof}

If the coefficients of $\widehat{\mb{H}}$ are independently generated from a continuous distribution, for example, $\mathcal{CN}(\mb{0}, \mb{I})$, the condition of \emph{general position} is satisfied with probability one.

In practice, the condition of general position may not be satisfied. Next, we relate the infinite SNR capacity to the rank of the channel $\mb{H}$.
\begin{corollary}
The infinite SNR capacity of MIMO channel satisfies
  \begin{equation} \label{eq:Rate_MIMO_rank}
  2 \mr{rank}(\mb{H}) \leq \overline{C}_{\mr{1bit, MIMO}}
    \leq \log_2 \left( K\left( \Nr, \mr{rank}(\mb{H}) \right) +1 \right),
  \end{equation}
  when $\Nr > \mr{rank}(\mb{H})$ and
  \begin{eqnarray}
    \overline{C}_{\mr{1bit, MIMO}} = 2 \Nr,
  \end{eqnarray}
  when $\Nr = \mr{rank}(\mb{H})$.
\end{corollary}
\begin{IEEEproof}
In the corollary, there are two different cases:
\begin{enumerate}
  \item $\Nr = \mr{rank}(\mb{H})$;
  \item $\Nr > \mr{rank}(\mb{H})$.
\end{enumerate}

In the first case, the set of rows of $\mb{H}$ satisfies the condition of general position by Definition 1. In addition, $\Nr= \mr{rank}(\mb{H})$ implies $\Nt \geq \Nr$. Thus, by Proposition 3, we have $\overline{C}_{\mr{1bit, MIMO}} = 2 \Nr$.

In the second case, without loss of generality, assume that the first $\mr{rank}(\mb{H})$ rows are linearly independent. Consider a channel matrix consisting of these rows. By the argument of the first case, the capacity of this channel is $2\mr{rank}(\mb{H})$. Therefore, $\overline{C}_{\mr{1bit, MIMO}} \geq 2\mr{rank}(\mb{H})$. For the upper bound, the subspace spanned by the columns of $\mb{H}$ has dimension of $\mr{rank}(\mb{H})$ (but may not in general position). By the dual statement of Lemma 2 and Proposition \ref{prop:MIMO}, the channel capacity has the upper bound $\log_2 \left( K\left( \Nr, \mr{rank}(\mb{H}) \right) +1 \right)$.
\end{IEEEproof}

The function $K(\Nr, \Nt)$ has the following properties:
\begin{enumerate}
    \item $K(1, \Nt) = 4$;
    \item $K(\Nr, 1)= 4\Nr$.
\end{enumerate}

Property 1 implies that the high SNR capacities of SISO and MISO channel are $\log_2 K(1, \Nt)=2$ bps/Hz.
Combining Property 2 and Proposition \ref{prop:MIMO}, we obtain that the infinite SNR capacity of SIMO channel is between $\log_2 (4 \Nr)$ and $\log_2 (4 \Nr +1 )$, which is \eqref{eq:Rate_SIMO} in Proposition \ref{prop:SIMO}.


\begin{figure}[t]
\begin{centering}
\includegraphics[scale=.6]{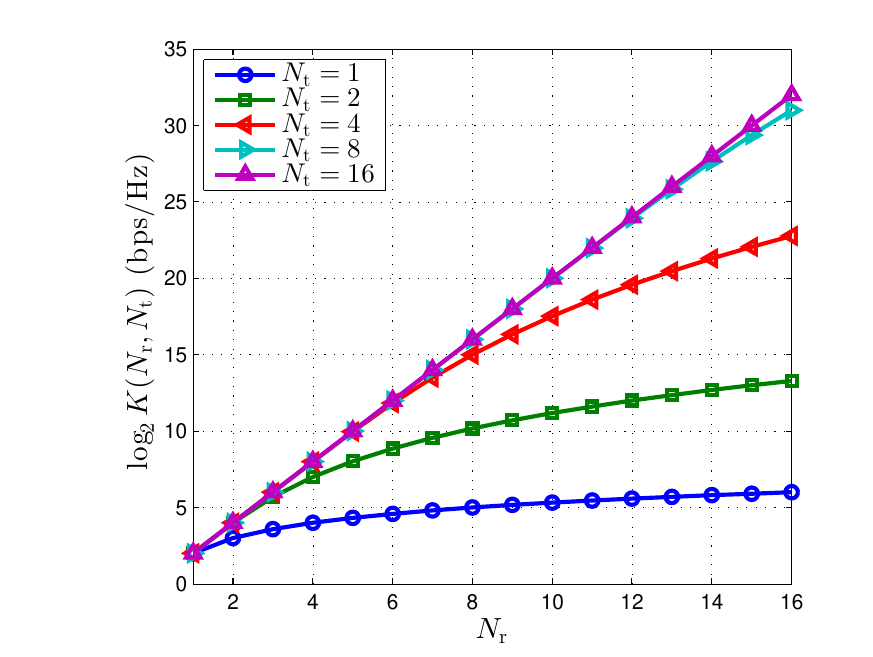}
\vspace{-0.1cm}
\centering
 \caption{$\log_2 K(\Nr, \Nt)$ versus $\Nr$ for different $\Nt$. $K(\Nr, \Nt)$ is defined as $2 \sum_{k=0}^{2\Nt-1} \binom{2\Nr-1}{k}$ and $\log_2 K(\Nr, \Nt)$ is close to the high SNR capacity of the MIMO channel with $\Nt$ transmit antennas and $\Nr$ receive antennas. } \label{fig:K_func_1}
\end{centering}
\vspace{-0.3cm}
\end{figure}

\begin{figure}[t]
\begin{centering}
\includegraphics[scale=.6]{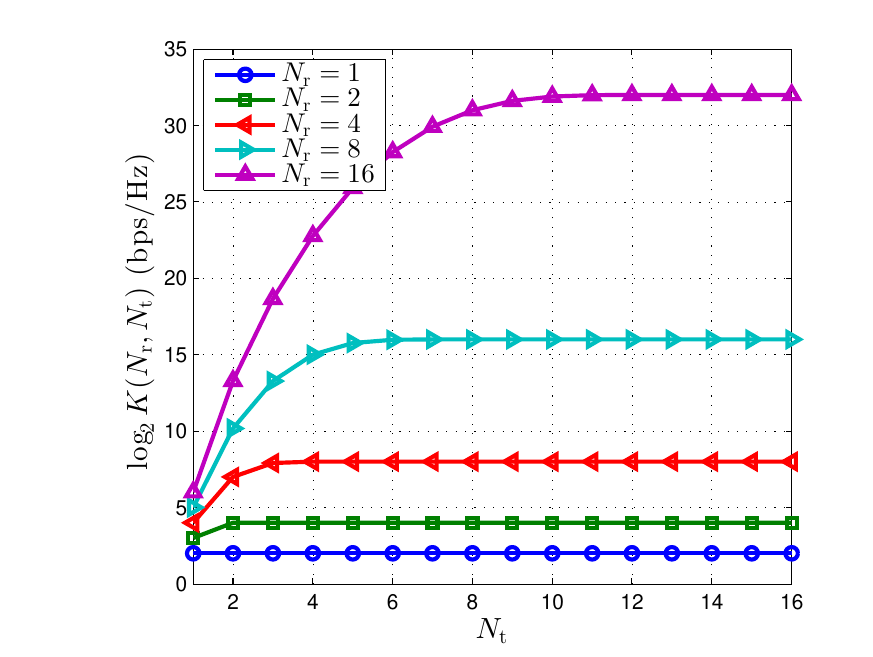}
\vspace{-0.1cm}
\centering
 \caption{$\log_2 K(\Nr, \Nt)$ versus $\Nt$ for different $\Nr$. $K(\Nr, \Nt)$ is defined as $2 \sum_{k=0}^{2\Nt-1} \binom{2\Nr-1}{k}$ and $\log_2 K(\Nr, \Nt)$ is close to the high SNR capacity of the MIMO channel with $\Nt$ transmit antennas and $\Nr$ receive antennas. }\label{fig:K_func_2}
\end{centering}
\vspace{-0.3cm}
\end{figure}

In Fig. \ref{fig:K_func_1} and Fig. \ref{fig:K_func_2}, we plot $\log_2 K(\Nr, \Nt)$, which is very close to the high SNR capacity $\overline{C}_{\mr{1bit, MIMO}}$ as shown in \eqref{eq:Rate_MIMO}. In Fig. \ref{fig:K_func_1}, $\log_2 K(\Nr, \Nt)$ is a strictly increasing function of $\Nr$.
In Fig. \ref{fig:K_func_2}, we see that $\log_2 K(\Nr, \Nt)$ increases fast with $\Nt$ and quickly becomes saturated when $\Nt \geq \frac{\Nr}{2}$. When $\Nt \geq \Nr$, $\log_2 K(\Nr, \Nt)= 2 \Nr$.
Moreover, we see that $K(m,n) \neq K(n, m)$ in general. This means that the capacity of a quantized $m \times n$ channel is different from that of a quantized $n \times m$ channel. This is in striking contrast with unquantized MIMO systems with \ac{CSIT}.


\section{Bounds of MIMO Channel Capacity with One-Bit Quantization at Finite SNR} \label{sec:finite_SNR}
In this section, we turn to the channel capacity at finite SNR. We will propose a new upper bound and discuss two lower bounds.

\subsection{Upper Bound of MIMO Channel Capacity at finite SNR}
\begin{proposition}
  The capacity of MIMO channel with one-bit quantization has the upper bound,
  \begin{eqnarray} \label{eq:MIMO_ub}
  C_{\mr{1bit}}^{\mr{ub}} = 2 \Nr \left(1 - \Hb \left( Q \left( \sqrt{ \frac{ \Pt \sigma_{\max}^2 }{\Nr} } \right) \right) \right),
\end{eqnarray}
where $\sigma_{\max}$ is the largest singular value of $\mb{H}$.
\end{proposition}
\begin{IEEEproof}
First, we have,
\begin{eqnarray}
  I(\mb{x}; \mb{r}) &=& \cH(\mb{r}) - \cH(\mb{r}|\mb{x}) \\
  & \stackrel{(a)}{\leq}& 2 \Nr - \cH(\mb{r}| \mb{x}),
\end{eqnarray}
where $(a)$ follows from that there are at most $2\Nr$ quantization outputs.

Next, we derive a lower bound of $\cH(\mb{r}|\mb{x})$.
For a given transmitted signal $\mb{x}=\mb{x}'$, denote $\mb{z}' = \mb{H} \mb{x}'$ and $\widehat{\mb{z}}' = [\Re(\mb{z}')^T, \Im(\mb{z}')^T]^T$. The conditional entropy of $\mb{r}$ given $\mb{x}=\mb{x}'$ is,
\begin{eqnarray}
  \cH( \mb{r}| \mb{x} = \mb{x}')
  &\stackrel{(a)}{=}& \sum_{j=1}^{\Nr} \cH( {\mb{r}}_j| {\mb{x}'}) \\
  &\stackrel{(b)}{=}& \sum_{j=1}^{2 \Nr} \Hb \left( Q \left( \sqrt{2} \widehat{\mb{z}}'_{j} \right) \right) \\
  &\stackrel{(c)}{=}& \sum_{j=1}^{2 \Nr} \Hb \left( Q \left( \sqrt{2 \left(\widehat{\mb{z}}'_{j} \right)^2} \right) \right),
\end{eqnarray}
where $(a)$ follows from that the noises across different antennas are independent, $(b)$ follows from that the in-phase and quadrature parts of the noises are independent and Gaussian distributed with variance $\frac{1}{2}$, $(c)$ follows from that $\Hb(Q(x))$ is an even function of $x$.
Next, note that
\begin{eqnarray}
  \sum_{j=1}^{2 \Nr} \left(\widehat{z}'_j \right)^2 = \|\mb{z}'\|^2 \leq \|\mb{x}'\|^2 \sigma_{\max}^2,
\end{eqnarray}
and $\Hb\left( Q\left( \sqrt{x}\right)\right)$ is decreasing and convex in $x$ (see \cite{Dabeer_SPAWC06} for the proof of convexity),
\begin{eqnarray}
  \cH( \mb{r}| \mb{x} = \mb{x}') &{\geq} & 2 \Nr \Hb \left( Q \left( \sqrt{ \frac{ \|\mb{x}'\|^2 \sigma_{\max}^2 }{\Nr} } \right) \right).
\end{eqnarray}

The conditional entropy of $\mb{r}$ is,
\begin{eqnarray}
  \cH({\mb{r}}|{\mb{x}}) = \mathbb{E}_{\mb{x}} \left[ 2 \Nr \Hb \left( Q \left( \sqrt{ \frac{ \|\mb{x}\|^2 \sigma_{\max}^2}{\Nr} } \right) \right) \right].
\end{eqnarray}

We want to minimize $\cH(\mb{r}|\mb{x})$ subject to the power constraint $\mathbb{E}\left[ \|\mb{x} \|^2 \right] \leq \Pt$.
Since $\Hb(Q(\sqrt{z}))$ is convex and decreasing in $z$,
\begin{eqnarray}
  \cH(\mb{r}|\mb{x}) &\geq & 2 \Nr \Hb \left( Q \left( \sqrt{ \frac{ \mathbb{E}[\|\mb{x}\|^2] \sigma_{\max}^2}{\Nr} } \right) \right) \\
  & \geq & 2 \Nr \Hb \left( Q \left( \sqrt{ \frac{ \Pt \sigma_{\max}^2 }{\Nr} } \right) \right).
\end{eqnarray}
Therefore, the upper bound in \eqref{eq:MIMO_ub} is obtained.
\end{IEEEproof}

The upper bound is achieved when $\mb{H}$ has $\Nr$ same singular values (or equivalently, $\mb{H}\mb{H}^*= \sigma_{\max}^2 \mb{I}$). The transmission strategy is the simple channel inversion which will be discussed later.

This bound is loose when $\bH$ has less than $\Nr$ nonzero singular values (or equivalently, $\mr{rank}(\bH)<\Nr$). For example, by Proposition \ref{prop:SIMO}, the SIMO channel capacity at infinite SNR is around $\log_2(4 \Nr)$. But the upper bound in \eqref{eq:MIMO_ub} approaches $2 \Nr$ at high SNR, which is larger than $\log_2 (4 \Nr)$ when $\Nr \geq 2$.

At low SNR, we obtain
\begin{eqnarray}
  C_{\mr{1bit}}^{\mr{ub}} = \frac{2}{\pi } \frac{\Pt \sigma_{\max}^2 }{\ln 2 } + o(\Pt).
\end{eqnarray}
The unquantized MIMO channel capacity with CSIT is $\frac{\Pt \sigma^2}{\ln 2} + o(\Pt)$ at low SNR. Therefore, the one-bit quantization results in \emph{at least} $10 \log_{10} \frac{\pi}{2} \approx 1.96$ dB power loss.

\subsection{Lower Bounds on MIMO Channel Capacity at Finite SNR}

\subsubsection{Channel Inversion} \label{sec:channel_inversion}
 When $\mb{H} \mb{H}^*$ is invertible, a simple transmission strategy is to use channel inversion (CI) precoding  and QPSK signaling \cite{Mezghani_WSA07}. The transmitted symbol is
\begin{eqnarray}
  \mb{x} = \sqrt{\frac{\Pt}{\mr{tr}\left( \left( \mb{H} \mb{H}^* \right)^{-1} \right)}} \mb{H}^* \left( \mb{H} \mb{H}^* \right)^{-1} \mb{s},
\end{eqnarray}
where $\mb{s} \in \mathbb{C}^{\Nr \times 1}$ is a vector with independent QPSK entries that satisfies $\mathbb{E}[\mb{s} \mb{s}^*] = \mb{I}_{\Nr \times \Nr}$.
The expected transmission power is mounted as $\mathbb{E}[\mb{x} \mb{x}^*]= \Pt$.
The output of the quantizer with this choice of precoder is
\begin{eqnarray}
  \mb{r} = \mr{sgn} \left( \sqrt{\frac{\Pt}{\mr{tr}\left( \left( \mb{H} \mb{H}^* \right)^{-1} \right)}} \mb{s} + \mb{n}\right).
\end{eqnarray}

The channel decomposes into $\Nr$ parallel one-bit quantized SISO channels with same channel gain.
According to Lemma \ref{lemma:SISO}, the achievable rate is
\begin{eqnarray} \label{eq:Rate_CI}
  R_{\mr{1bit}}^{\mr{CI}} = 2 \Nr \left(1 - \Hb\left(  Q\left(\sqrt{\frac{\Pt}{\mr{tr}\left( \left( \mb{H} \mb{H}^* \right)^{-1} \right)}} \right) \right)\right).
\end{eqnarray}
If $\mb{H}\mb{H}^*= \sigma_{\max}^2 \mb{I}$, we find that $R_{\mr{1bit}}^{\mr{CI}}$ is equal to $C_{\mr{1bit}}^{\mr{ub}}$. This implies that the channel inversion transmission strategy is \emph{capacity-achieving} when the channel $\bH$ has $\Nr$ identical singular values.

Denote the eigenvalues of $\mb{H} \mb{H}^*$ as $\lambda_1 \geq \lambda_2 \geq \cdots \geq \lambda_{\Nr} > 0$. We have
\begin{eqnarray}
  \frac{\Pt}{\mr{tr}\left( \left( \mb{H} \mb{H}^* \right)^{-1} \right)} &=& \frac{\Pt}{\frac{1}{\lambda_1} + \frac{1}{\lambda_2}+ \cdots + \frac{1}{\lambda_{\Nr}}} \\
  & \geq & \frac{\Pt \lambda_{\Nr}}{\Nr} \\
  & = & \frac{\Pt \lambda_{1}}{\Nr} \cdot \frac{\lambda_{\Nr}}{\lambda_1}.
\end{eqnarray}
Note that $\lambda_1= \sigma_{\max}^2$, we conclude that the power loss of the channel inversion strategy compared to the optimum is at most $\left( 10\log_{10} \frac{\lambda_1}{\lambda_{\Nr}} \right)$ dB. Note that $\frac{ \sqrt{\lambda_1}}{ \sqrt{\lambda_{\Nr}}}$ is the condition number of the matrix $\bH$. Therefore, the channel inversion strategy is nearly optimum when the channel matrix has a small condition number.

At low SNR, we have
\begin{eqnarray}
  R_{\mr{1bit}}^{\mr{CI}}
  &=&\frac{2}{\pi } \frac{\Nr \Pt}{ \mr{tr}\left( \left( \mb{H} \mb{H}^* \right)^{-1} \right) \ln 2 } + o(\Pt) \\
  &=&\frac{2}{\pi } \frac{\Nr}{ \frac{1}{\lambda_1} + \frac{1}{\lambda_2} + \cdots \frac{1}{\lambda_{\Nr}}} \frac{P_{\mr{t}}} {\ln 2 } + o(P_{\mr{t}}).
\end{eqnarray}

In \cite[Theorem 2]{Mezghani_ISIT07}, the achievable rate of MIMO channel with independent QPSK signaling and one-bit ADCs at low SNR is shown to be
\begin{eqnarray}
  R_{\mr{1bit}}^\mr{QPSK}
  &=& \frac{2}{\pi} \mr{tr} \left(\mb{H} \mb{H}^* \right) \frac{P_{\mr{t}}}{\Nt \ln 2} + o(\Pt) \label{eq:QPSK_1bit_MIMO}\\
  &=& \frac{2}{\pi}  \frac{\lambda_1 + \lambda_2 + \cdots + \lambda_{\Nr}}{\Nt} \frac{P_{\mr{t}}}{ \ln 2} + o(\Pt).
\end{eqnarray}
When $\Nr=\Nt$, $R_{\mr{1bit}}^\mr{QPSK} \geq R_{\mr{1bit}}^{\mr{CI}}$ because of Jensen's inequality $\frac{1}{\mathbb{E}(\lambda)} \leq \mathbb{E}\left( \frac{1}{x}\right)$. This means that with relative small transmitter antenna array, the channel inversion method does not provide gain over the simple QPSK signaling. However, if $\Nt \gg \Nr$, then $R_{\mr{1bit}}^\mr{QPSK} < R_{\mr{1bit}}^{\mr{CI}}$. The reason is that the channel inversion strategy has array gain which increases with the number of transmitter antennas.

\subsubsection{Additive Quantization Noise Model (AQNM)}
It is known that Gaussian distributed noise minimises the mutual information \cite{Cover_Book12}. A lower bound of the capacity can be derived by assuming the quantization error as Gaussian distributed noise. Such a lower bound was given in \cite{Bai_Qing_ETT14, Mezghani_WSA07} as
\begin{eqnarray} \label{eq:AQNM}
  & & R^{\mr{AQNM}} \nonumber \\
  &=& \log_2 \left| \mr{I}_{\Nt} + \frac{\Pt}{\Nt} \mb{H}^* \mr{diag}\left\{ \frac{ 1 - \rho }{1+ \rho \frac{\Pt ||\mb{h}_i||^2}{\Nt}}\right\}_{i=1}^{\Nr} \mb{H}\right|,
\end{eqnarray}
where $\mb{h}_i^*$ is the $i$-th row of $\mb{H}$.
As pointed out in \cite{Mezghani_WSA07, Bai_Qing_ETT14, Orhan_ITA15}, this lower bound is quite tight at low SNR when the additive white Gaussian noise is dominating. As shown below, however, we find that this lower bound is loose at high SNR when the quantization noise is dominating.
\begin{eqnarray*}
  & & R^{\mr{AQNM}} \\
  & \leq & \log_2 \left| \mr{I}_{\Nt} +  \mb{H}^* \mr{diag}\left\{ \frac{ 1 - \rho}{\rho \
  \left\|\mb{h}_i \right\|^2 }\right\}_{i=1}^{\Nr} \mb{H}\right| \\
  & = & \log_2 \left| \mr{I}_{\Nt} +  \frac{1 - \rho}{\rho} \mb{H}^* \mr{diag}\left\{ \frac{1}{ ||\mb{h}_i|| }\right\} \mr{diag}\left\{ \frac{1}{ ||\mb{h}_i|| }\right\} \mb{H}\right| \\
  & = & \log_2 \left| \mr{I}_{\Nt} +  \frac{1 - \rho}{\rho} \widetilde{\mb{H}}^*  \widetilde{\mb{H}} \right| \\
  & = & \log_2 \left| \mr{I}_{\Nr} +  \frac{1 - \rho}{\rho} \widetilde{\mb{H}} \widetilde{\mb{H}}^*   \right| \\
  & = & \sum_i \log_2 \left(1 + \frac{1-\rho}{\rho} \lambda_i \left( \widetilde{\mb{H}} \widetilde{\mb{H}}^* \right)\right)
\end{eqnarray*}
where $\rho$ is the distortion factor (see Table 1 in \cite{Bai_Qing_ETT14} for the value of $\rho$), $\widetilde{\mb{H}}$ is obtained by normalizing each row of $\mb{H}$, and $\lambda_i \left(\widetilde{\mb{H}} \widetilde{\mb{H}}^*  \right)$ is the eigenvalue of the matrix $\widetilde{\mb{H}} \widetilde{\mb{H}}^* $.

Since each row of $\widetilde{\bH}$ has unit norm, then
\begin{eqnarray}
  \sum_i \lambda_i \left( \widetilde{\bH}\widetilde{\bH}^* \right)= \mr{tr} \left( \widetilde{\bH}\widetilde{\bH}^* \right) = \Nr.
\end{eqnarray}
Therefore,
\begin{eqnarray}
R^{\mr{AQNM}} & \stackrel{(a)}{\leq} & \Nr \log_2 \left(1+ \frac{1-\rho}{\rho}\right) \\
  & = & \Nr \log_2 \frac{1}{\rho},
\end{eqnarray}
where $(a)$ follows from that $\log_2(1+x)$ is concave in $x$. For one-bit quantization, the distortion factor
$\rho$ is $\frac{\pi-2}{\pi}$.
Therefore, we have,
\begin{eqnarray} \label{eq:AQNM_ub}
    & & R_{\mr{1bit}}^{\mr{AQNM}} \leq \Nr \log_2 \frac{\pi}{\pi-2} \approx 1.46 \Nr.
\end{eqnarray}
If $\bH \bH^{*}$ is invertible, we know that the achievable rate of channel inversion strategy in \eqref{eq:Rate_CI} approaches $2 \Nr$ at high SNR. Therefore, we conclude that this AQNM lower bound is loose at high SNR.


\section{Numerical Input Optimization Methods for the MIMO Channel}\label{sec:convex}


The channel inversion strategy only applies when $\mb{H} \mb{H}^*$ is invertible (or $\Nt \geq \Nr$ when $\mb{H}$ has IID Gaussian entries).
In this section, we propose a heuristic method that achieves the high SNR capacity $\overline{C}_{\mr{1bit, MIMO}}$ without requiring $\mb{H} \mb{H}^*$ being invertible. The basic idea is: for each possible quantization output $\mb{r}$, find the input signal $\mb{x}$ such that $\mb{r}=\mr{sgn}(\mb{H} \mb{x})$.

The input signals $\mb{x}$ are obtained by solving the following optimization problem,
\begin{subequations}\label{eq_benchmark_complex}
\begin{eqnarray}
  \textbf{P1:} & \mathop{\max}\limits_{\mb{x}} & d \\
   &\text{s.t.} &   \Re\left(  \mb{H} \mb{x}  \right) \odot \Re(\mb{r})  \geq d \cdot \mb{1}_{\Nr}, \label{eq:cons_sgn_real}\\
   & & \Im \left( \mb{H} \mb{x} \right) \odot \Im(\mb{r})  \geq d \cdot \mb{1}_{\Nr}, \label{eq:cons_sgn_imag}\\
   & &  \mb{x}^* \mb{x} \leq \Pt.
\end{eqnarray}
\end{subequations}
where the inequalities $\geq$ in \eqref{eq:cons_sgn_real} and \eqref{eq:cons_sgn_imag} are applied componentwise.
The objective is to maximize the minimum distance between the the noiseless received signal $\mb{Hx}$ and the threshold of the one-bit ADCs, which is zero. If $d \geq 0$, then \eqref{eq:cons_sgn_real} and \eqref{eq:cons_sgn_imag} imply $\mr{sgn} (\mb{H} \mb{x}) = \mb{r}$.

Using the notation given in \eqref{eq:notation_real}, we rewrite Problem \textbf{P1} in a compact form as,
\begin{subequations}\label{eq_benchmark_real}
\begin{eqnarray}
  \textbf{P2:} & \mathop{\max}\limits_{\widehat{\mb{x}}} & d \\
   &\text{s.t.} &  \left(\mr{diag}(\widehat{\mb{r}}) \widehat{\mb{H}} \right) \widehat{\mb{x}} \geq d \cdot \mb{1}_{2\Nr}, \label{eq:cons_sgn} \\
   & &  \widehat{\mb{x}}^T \widehat{\mb{x}} \leq \Pt.
\end{eqnarray}
\end{subequations}

For fixed $\widehat{\mb{r}}$, the inequality constraint \eqref{eq:cons_sgn} is linear and thus convex. Therefore, the problem $\textbf{P2}$ is convex and can be solved by software solver such as CVX  \cite{Boyd_04}.

There are a total of $2^{2\Nr}$ possible values of $\hat{\mb{r}}$ and thus $2^{2\Nr}$ different optimization problems.
Denote the optimal value of Problem \textbf{P2} as $d^{\star}(\widehat{\mb{r}})$.
Note that if $\widehat{ \mb{x} } = \mb{0}$, the value of the objective function in Problem \textbf{P2} is zero regardless of $\widehat{\mb{r}}$.
Therefore, a lower bound of  $d^{\star}(\widehat{\mb{r}})$ is zero.
When $d^{\star}(\widehat{\br})>0$, the corresponding $\widehat{\bx}$ is put in the transmitter constellation.
There may be many optimization problems with $d^{\star}(\widehat{\mb{r}})=0$, for example in the SIMO channel, $4 \Nr$ out of the $2^{2 \Nr}$ problems has objective value $d^{\star}(\widehat{\br})>0$.
When $\Nr$ is large, it is inefficient to solve each of the convex problem.

We now provide a method to reduce the complexity.
%
%
%
We can see that $d^{\star}(\widehat{\mb{r}})>0$ if and only if there exists $\widehat{\mb{x}}$ satisfying
\begin{equation} \label{eq:feasi_cond}
  \left(\mr{diag}(\widehat{\mb{r}}) \widehat{\mb{H}} \right) \widehat{\mb{x}} > \mb{0}.
\end{equation}
Therefore, it is efficient to check the feasibility of \eqref{eq:feasi_cond} before solving the optimization problem.  Actually, \eqref{eq:feasi_cond} is a system of linear inequalities and can be solved by a methd called `Fourier-Dines-Motzkin Elimination' \cite{Dines_Math19, Dantzig_73} (see \cite{Dines_Math19, Dantzig_73} for more details of the method).


The transmitter constellation is composed of the nonzero solutions of the $2^{2\Nr}$ convex optimization problems and the zero symbol. To reduce the PAPR, the zero symbol is often not included in the constellation. Therefore, in our simulations, the transmitter constellation only contains the nonzero solutions of the convex optimization problems.
 Instead of transmitting the symbols with equal probability, we can also optimize the probabilities of each symbol using the well-known Blahut-Arimoto algorithm \cite{Blahut_IT72}. The performance improvement is shown in our simulation results.
We find that optimization of the transmission probability of the symbols only provides small gain over a uniform distribution in the low and medium SNR regime.

In the appendix, we prove that a lower bound of the achievable rate of this heuristic method is
\begin{eqnarray} \label{eq:CV_lb}
 R_{\mr{1bit}}^{ \mr{CO, lb}} = \min\{a_1, a_2\} - 2 \Nr \Hb\left( Q \left( \sqrt{ \alpha \left(\mb{H} \right) \Pt} \right)\right),
\end{eqnarray}
where $\alpha(\bH)$ is a constant depending on the channel $\bH$ and
\begin{eqnarray*}
  a_1&=&-(M-1) \frac{q}{M}\log_2 \frac{q}{M} \\
  & & \quad -\left(1- \frac{(M-1)q}{M} \right) \log_2 \left(1- \frac{(M-1)q}{M} \right), \\
  a_2&=&-q\log_2 \frac{q}{M}-\left(1-q \right) \log_2 \left(1- q \right),
\end{eqnarray*}
where $M$ is number of the convex optimization with nonzero objective function value and $q = \left(1-Q\left(\sqrt{ \alpha(\bH) \Pt}\right) \right)^{2\Nr}$.

At high SNR, this lower bound converges to $M$. Note that when the condition of general position is satisfied, $M=K(\Nr, \Nt)$. This implies that at high SNR, the rate of this convex optimization based method approaches the infinite SNR capacity.

\section{MmWave Channel with One-Bit Quantization}
In mmWave communications, the channel matrix $\mb{H}$ is usually assumed to be low rank due to sparse scattering in the channel \cite{Daniels_Book14} and therefore does not satisfy the strict condition of \emph{general position} in Proposition \ref{prop:MIMO}. As a result, we cannot directly obtain the infinite SNR capacity of the mmWave channel.

In this section, the mmWave MIMO channel is modeled using a ray-based model with $L$ paths. We also assume that uniform planar arrays (UPA) in the $yz$-plane are deployed at the transmitter and receiver. Denote $\alpha_{\ell}$, $\varphi_{\mr{r}\ell}$ $(\theta_{\mr{r}\ell})$, $\varphi_{\mr{t}\ell}$ $(\theta_{\mr{t}\ell})$ as the strengths, the azimuth (elevation) angles of arrival and the angle of departure of the $\ell$th path, respectively.  The array response vectors at the transmitter or receiver is given by \cite{Balanis_Book12}
\begin{eqnarray}
  \mathbf{a}(\varphi, \theta) &=& \frac{1}{\sqrt{N}}[1, e^{\j k ( m \sin(\varphi) \sin(\theta) + n \cos(\theta))} , ..., \nonumber \\
  & & e^{\j k ((Y-1) \sin(\varphi) \sin(\theta) + (Z-1) \cos(\theta))}]^T,
\end{eqnarray}
where $0 \leq m < Y$ and $0 \leq n <Z$ are the $y$ and $z$ indices of an antenna element respectively. Herein, $k= \frac{2\pi}{\lambda} d$ where $\lambda$ is the wavelength and $d$ is the inter-element spacing.
Hence, the channel matrix is,
  \begin{eqnarray}
    \mathbf{H} &=& \sum_{\ell=1}^{L} \alpha_{\ell} \mathbf{a}_{\mr{r}}(\varphi_{\mr{r}\ell}, \theta_{\mr{r}\ell}) \mathbf{a}_{\mr{t}}^*(\varphi_{\mr{t}\ell},
    \theta_{\mr{t}\ell}) \\
               &=& \mb{A}_{\mr{r}} \mb{\Sigma} \mb{A}_{\mr{t}}^* \label{eq:MIMO_ULA}
  \end{eqnarray}
  where
  $\mb{A}_{\mr{r}} = \left[\mb{a}_{\mr{r}}(\varphi_{\mr{r}1}, \theta_{\mr{r}1}) , \mb{a}_{\mr{r}}(\varphi_{\mr{r}2}, \theta_{\mr{r}2}), \cdots, \mb{a}_{\mr{r}}(\varphi_{\mr{r}L}, \theta_{\mr{r}L})\right]$,
  $\mb{\Sigma} = \mr{diag}(\alpha_1, \alpha_2, \cdots, \alpha_L)$ and $\mb{A}_{\mr{t}} = \left[\mb{a}_{\mr{t}}(\varphi_{\mr{t}1}, \theta_{\mr{t}1}) , \mb{a}_{\mr{t}}(\varphi_{\mr{t}2}, \theta_{\mr{t}2}), \cdots \mb{a}_{\mr{t}}(\varphi_{\mr{t}L}, \theta_{\mr{t}L})\right]$.

The number of multipaths tends to be lower in the mmWave band compared with lower frequencies.
Meanwhile, large antenna arrays are usually deployed to obtain array gain for combatting the path loss. Hence, we suppose that $L<\min\{\Nt, \Nr\}$ in mmWave MIMO channels.

\begin{corollary}\label{coro:MIMO_ULA}
  For a mmWave MIMO system with $L \left(L< \min (\Nr, \Nt)\right)$ paths and one-bit quantization, the infinite SNR capacity satisfies,
  \begin{eqnarray}
  \log_2 \left( K(\Nr, L) \right) \leq \overline{C}_{\mr{1bit, mmW}} \leq \log_2 \left( K(\Nr, L) +1 \right).
  \end{eqnarray}
\end{corollary}
\begin{IEEEproof}
First denote
\begin{eqnarray}
    \widehat{\mb{A}}_{\mr{r}} = \left(
                                  \begin{array}{cc}
                                    \Re(\mb{A}_{\mr{r}})  & -\Im (\mb{A}_{\mr{r}})  \\
                                    \Im(\mb{A}_{\mr{r}}) & \Re(\mb{A}_{\mr{r}}) \\
                                  \end{array}
                                \right).
\end{eqnarray}
When $L \leq \Nr$, any $2L \times 2L$ submatrix of $\widehat{\mb{A}}_{\mr{r}}$
has full rank with probability one since $\varphi_{\mr{r}\ell}$ and $\theta_{\mr{r}\ell}$ are generated from continuous distribution. Therefore, $\mb{A}_{\mr{r}}$ satisfies the condition of general position.

In addition, $\mb{A}_{\mr{t}}$ has rank $L$ with probability one and thus $\{\mb{A}_{\mr{t}}^* \mb{x}: \mb{x} \in \mathbb{C}^{\Nt \times 1}\}$ represents the $L$-dimensional complex space. Therefore, the channel is equivalent to a $\Nr \times L$ channel satisfying the condition of general position.
Combining these with Proposition \ref{prop:MIMO}, we obtain Corollary \ref{coro:MIMO_ULA}.
\end{IEEEproof}
Corollary \ref{coro:MIMO_ULA} shows that multipath is helpful in improving the infinite SNR capacity in mmWave MIMO systems.

As $\mb{H} \mb{H}^*$ is not invertible in mmWave systems, the channel inversion approach cannot be used to design the transmitted symbols.
Instead, the convex optimization approach proposed in Section \ref{sec:convex} has to be used.

  Next, to provide intuition, we consider a mmWave MIMO channel with only one path and propose a capacity-achieving transmission strategy. If $L=1$, the channel in \eqref{eq:MIMO_ULA} degenerates to
  \begin{eqnarray}
    \mb{H}= \alpha \mb{a}_{\mr{r}}(\varphi_{\mr{r}}, \theta_{\mr{r}}) \mb{a}_{\mr{t}}^*(\varphi_{\mr{t}}, \theta_{\mr{t}}).
  \end{eqnarray}
%
  Following the same logic in the MISO setting, matched filter beamforming is used at the transmitter to obtain the array gain. The resulting channel is equivalent to a SIMO channel with channel coefficients as $\mb{h} = \alpha ||\mb{a}_{\mr{t}}(\varphi_{\mr{t}}, \theta_{\mr{t}})||^2 \mb{a}_{\mr{r}}(\varphi_{\mr{r}}, \theta_{\mr{r}})$. Then the cutting plane method (details of the method are presented in the next section) can be used to design the transmitted symbols. Therefore, the transmitted symbols will be
  \begin{eqnarray}
    \mb{x} = \mb{a}_{\mr{t}} (\varphi_{\mr{t}}, \theta_{\mr{t}}) s
  \end{eqnarray}
  where $s$ is the symbol obtained by the cutting plane method in the equivalent SIMO channel.

\section{Simulation Results}
In this section, we first present simulation results of the capacities or achievable rates of the SIMO, MISO and the MIMO channel at high and low SNR. The performance of the proposed convex optimization method will also be shown. Last we consider the mmWave channel model and evaluate the achievable rates.
\subsection{MISO Channel with One-Bit Quantization}

\begin{figure}[t]
\begin{centering}
\includegraphics[scale=.6]{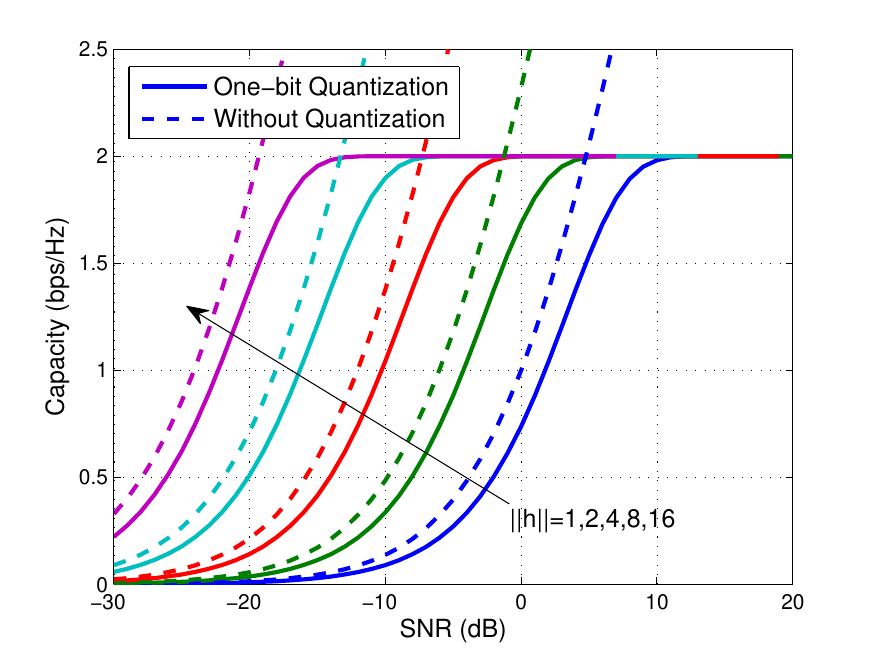}
\vspace{-0.1cm}
\centering
 \caption{Comparison of MISO channel capacity with one-bit quantization and that without quantization for different $||\mb{h}||$'s.}\label{fig:MISO_capacity}
\end{centering}
\vspace{-0.5cm}
\end{figure}

In Fig. \ref{fig:MISO_capacity}, we plot the MISO channel capacity with and without one-bit quantization for different values of $||\mb{h}||$. First, we see that the channel capacity with one-bit quantization approaches $2$ bps/Hz in the high SNR regime. Second, the transmitter antenna array provides power gain as shown in the figure. There is about $2$ dB power loss in the low and medium SNR regimes, which verifies our analysis in \eqref{eq:MISO_approx}. Third, when $||h||=16$, the capacity is close to the upper bound when SNR is larger than $-15$ dB. We see that the ``high SNR'' in our analysis can be very low in the practice thanks to the array gain provided by the multiple antennas.

\subsection{SIMO Channel with One-Bit Quantization}

\begin{figure}[t]
\begin{centering}
\includegraphics[scale=.6]{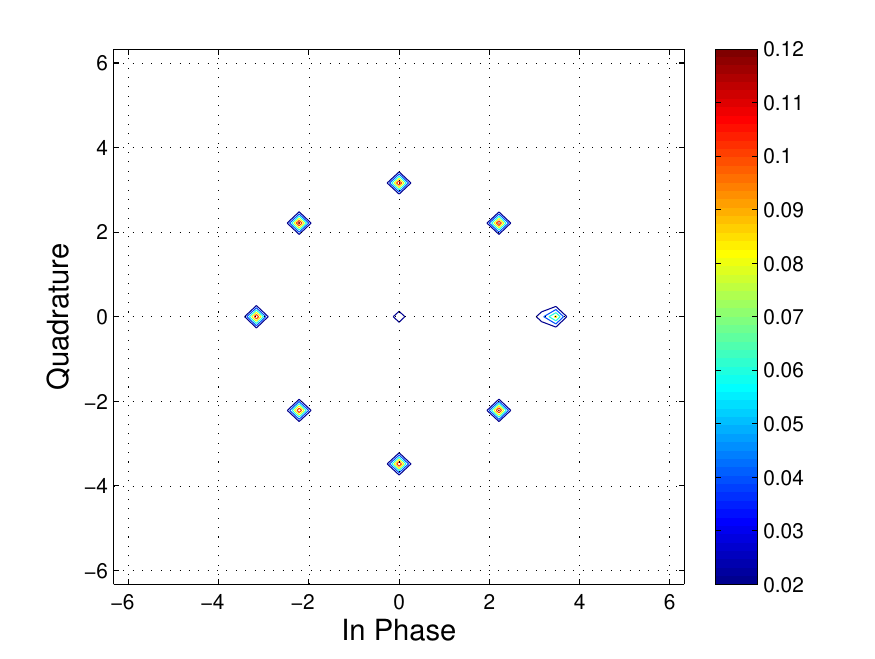}
\vspace{-0.1cm}
\centering
 \caption{The optimal input distribution of the SIMO channel where $\mb{h}=[e^{\j \pi/8}, e^{-\j \pi/8}]^T$. The transmission power $\Pt = 10$ and the achieved rate is about $2.52$ bps/Hz.}\label{fig:Input_constellation}
\end{centering}
\vspace{-0.3cm}
\end{figure}

\begin{figure}[t]
\begin{centering}
\includegraphics[scale=.6]{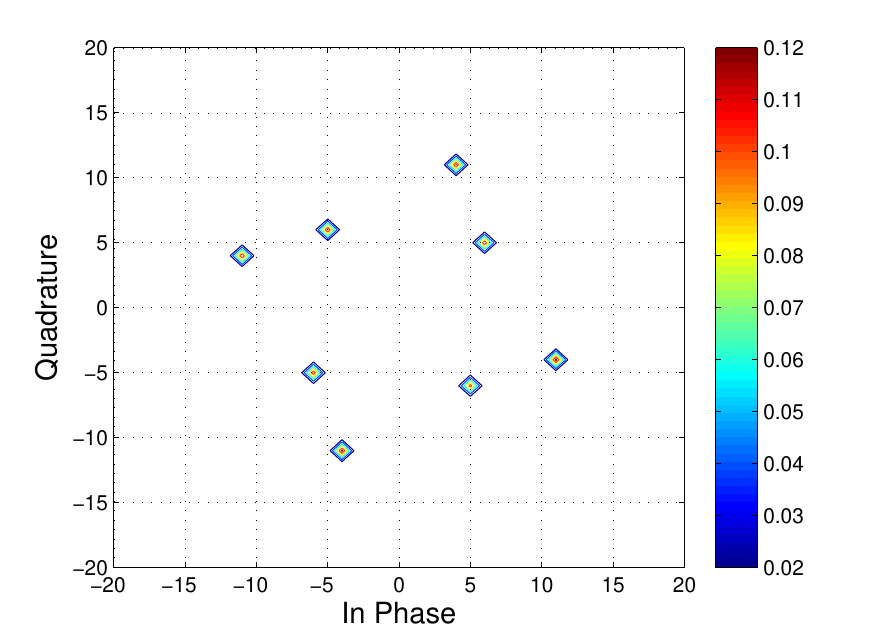}
\vspace{-0.1cm}
\centering
 \caption{The optimal input distribution of the SIMO channel where $\mb{h}=[1 \; 2e^{-j \pi/3}]^T$. The transmission power $\Pt=20$ dB and the achievable rate is $3.0050$ bps/Hz.}\label{fig:N_2_input_contour_nonregular}
\end{centering}
\vspace{-0.3cm}
\end{figure}

In the SIMO channel, we can obtain the capacity-achieving input distribution using the cutting plane method proposed in \cite[Sec. IV-A]{Huang_Jianyi_IT05} and used in \cite{Krone_SARNOFF12, Singh_TCOM09}. For this method, we take a fine quantized discrete grid on the region, for example, $\{x: -3\sqrt{\Pt}\leq \Re(x) \leq 3\sqrt{\Pt}, -3\sqrt{\Pt}\leq \Im(x) \leq 3\sqrt{\Pt}\}$, as the possible inputs and optimize their probabilities. The algorithm is an iterative algorithm which converges quite fast in several tens of iterations. Each iteration is a convex problem which can be solved efficiently.


In Fig. \ref{fig:Input_constellation}, we show a simple case when $\mb{h}=[e^{\mr{j} \pi/8} , e^{-\mr{j} \pi/8}]^T$. It is interesting to find that the optimal input constellation contains the rotated 8-PSK symbols and the symbol zero. For other general channels, the optimal constellation may not be regular. For example, in Fig. \ref{fig:N_2_input_contour_nonregular}, we show the optimal input distribution when $\mb{h}= [1 , 2e^{j \pi/3}]^T$.


\subsection{MIMO Channel with One-Bit Quantization}

\begin{figure}[t]
\begin{centering}
\includegraphics[scale=.65]{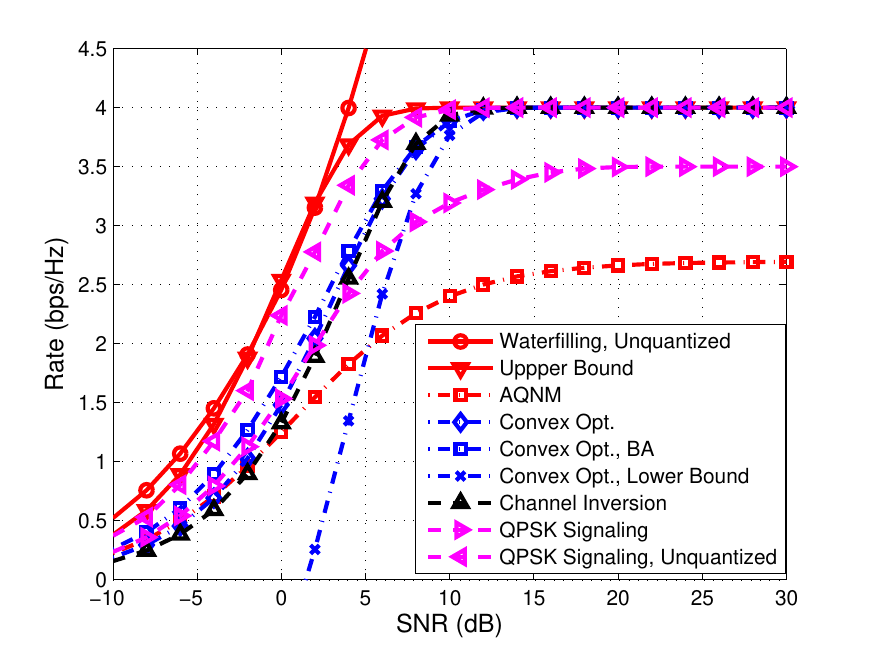}
\vspace{-0.1cm}
\centering
 \caption{Achievable rate of the $2 \times 2$ MIMO system with one-bit ADCs in the medium and high SNR regimes. }\label{fig:Rate_Nr_2_Nt_2}
\end{centering}
\vspace{-0.3cm}
\end{figure}

\begin{figure}[t]
\begin{centering}
\includegraphics[scale=.65]{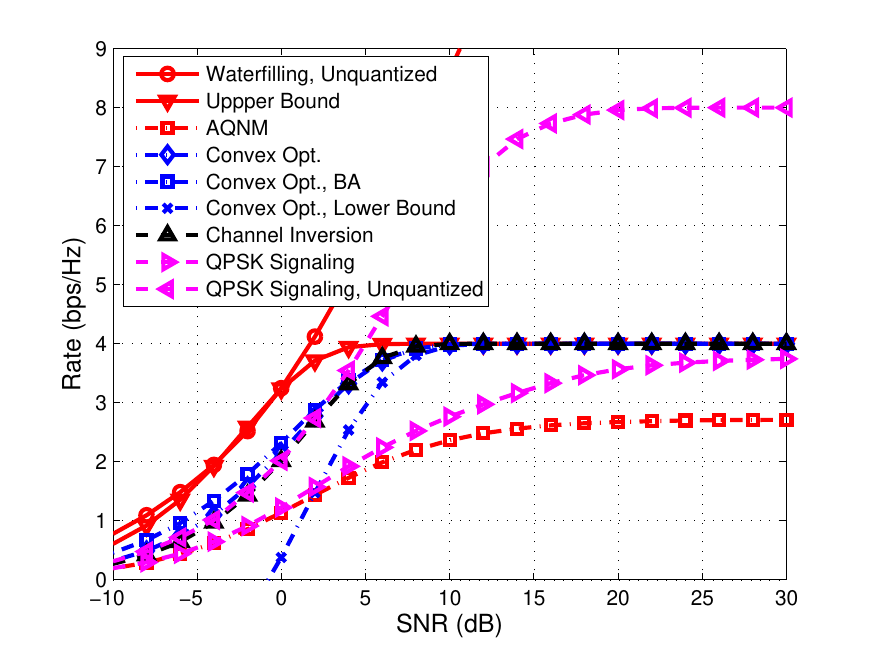}
\vspace{-0.1cm}
\centering
 \caption{Achievable rate of the $2 \times 4$ MIMO system with one-bit ADCs in the medium and high SNR regimes. }\label{fig:Rate_Nr_2_Nt_4}
\end{centering}
\vspace{-0.3cm}
\end{figure}

\begin{figure}[t]
\begin{centering}
\includegraphics[scale=.65]{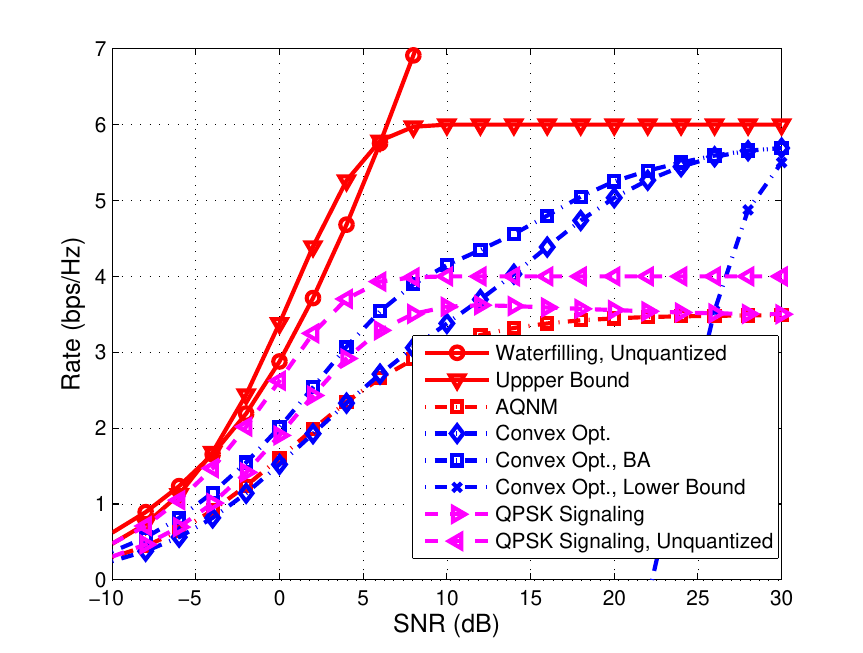}
\vspace{-0.1cm}
\centering
 \caption{Achievable rate of the $3 \times 2$ MIMO system with one-bit ADCs in the medium and high SNR regimes. }\label{fig:Rate_Nr_3_Nt_2}
\end{centering}
\vspace{-0.3cm}
\end{figure}

\begin{figure}[t]
\begin{centering}
\includegraphics[scale=.65]{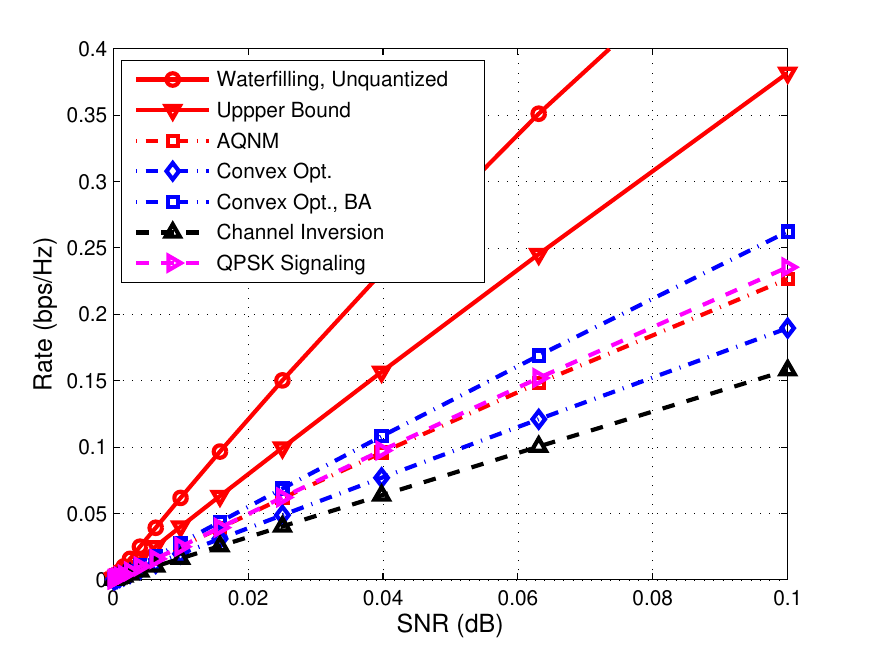}
\vspace{-0.1cm}
\centering
 \caption{Achievable rate of the $2 \times 2$ MIMO system with one-bit ADCs in the low SNR regime (below $-10$ dB). Note the SNR is in linear scale.}\label{fig:Rate_Nr_2_Nt_2_low_SNR}
\end{centering}
\vspace{-0.3cm}
\end{figure}

\begin{figure}[t]
\begin{centering}
\includegraphics[scale=.65]{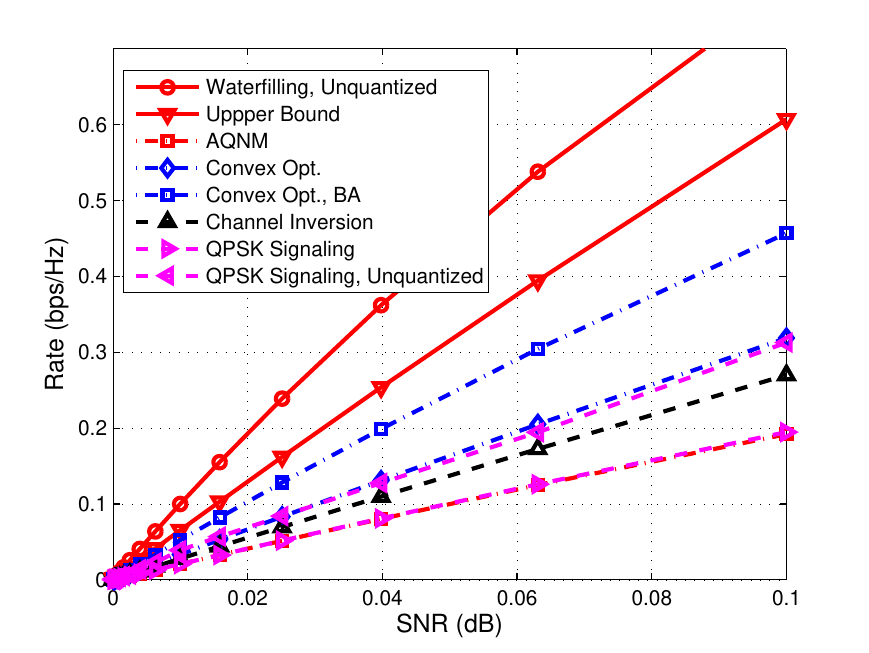}
\vspace{-0.1cm}
\centering
 \caption{Achievable rate of the $2 \times 4$ MIMO system with one-bit ADCs in the low SNR regime (below $-10$ dB). Note the SNR is in linear scale. }\label{fig:Rate_Nr_2_Nt_4_low_SNR}
\end{centering}
\vspace{-0.3cm}
\end{figure}

\begin{figure}[t]
\begin{centering}
\includegraphics[scale=.65]{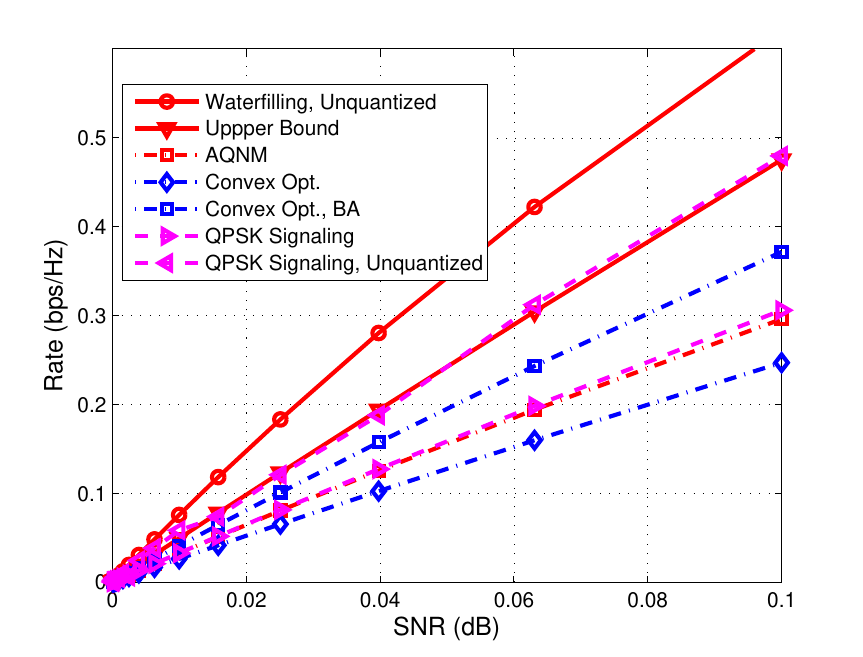}
\vspace{-0.1cm}
\centering
 \caption{Achievable rate of the $3 \times 2$ MIMO system with one-bit ADCs in the low SNR regime (below $-10$ dB). Note the SNR is in linear scale. }\label{fig:Rate_Nr_3_Nt_2_low_SNR}
\end{centering}
\vspace{-0.3cm}
\end{figure}

In this part, we illustrate the average achievable rates of MIMO system with one-bit quantization.
The channel coefficients are generated from $\mathcal{CN}(0, 1)$ distribution independently.
The input alphabet, which contains $2^{2\Nr}$ input symbols, are constructed by two methods, i.e., channel inversion method and convex optimization method.
For the convex optimization method, these symbols are transmitted with equal probabilities $2^{-2 \Nr}$ or with probabilities optimized by the Blahut-Arimoto algorithm (denoted as ``BA'' in the figures).

In Figs. \ref{fig:Rate_Nr_2_Nt_2}, \ref{fig:Rate_Nr_2_Nt_4} and \ref{fig:Rate_Nr_3_Nt_2}, we plot the achievable rates in the medium and high SNR regimes when $\Nr \times \Nt$ is $2 \times 2$, $2 \times 4$ and $3 \times 2$, respectively. Note that the channel inversion method can be used in the first two cases but not in the third case.

First, the achievable rates of convex optimization method, denoted as `Convex Opt.', converge to the upper bound $2 \Nr=4$ bps/Hz in Fig. \ref{fig:Rate_Nr_2_Nt_2} and Fig. \ref{fig:Rate_Nr_2_Nt_4} and $\log_2 K(3, 2)= \log_2 52 \approx 5.7$ bps/Hz in Fig. \ref{fig:Rate_Nr_3_Nt_2}. This verifies that the convex optimization method can approach the infinite SNR capacity in Proposition \ref{prop:MIMO}.
Its lower bound given in \eqref{eq:CV_lb} denoted as `Convex Opt., Lower Bound' is also plotted in these figures. We can see that the bound is tight at high SNR.

Second, let us examine the bounds at finite SNR. The theoretical upper bound given in \eqref{eq:MIMO_ub} is quite tight in Fig. \ref{fig:Rate_Nr_2_Nt_2} and Fig. \ref{fig:Rate_Nr_2_Nt_4} but loose in Fig. \ref{fig:Rate_Nr_3_Nt_2}. The reason is that the channel in Fig. \ref{fig:Rate_Nr_3_Nt_2} does not has full row rank. Namely, the rank of the channel is 2, while there are $\Nr=3$ receiver antennas. The channel inversion method works well at high SNR but has worse performance in the medium and low SNR regimes compared to the convex optimization method. But when $\Nr=4$, the gap between the performances of these two methods is negligible. As expected, the AQNM lower bound given in \eqref{eq:AQNM_ub} is loose at high SNR in all three figures.

Third, the rates of the channel without quantization are computed using the usual SVD precoding and waterfilling approach. We can see that the power loss of the quantized systems compared to the unquantized systems is less than $5$ dB at medium SNR.

Last, let us evaluate the performance of independent QPSK signaling across the transmitter antennas. For the case of independent QPSK signaling and one-bit quantization, we find that although its rate is larger than the AQNM lower bound, its performance is worse than the two proposed methods at high SNR.
We also simulate the case of QPSK signaling without quantization where the input is discrete and the output is continuous (see for example \cite{Baccarelli_JSAC01} for the computation of the rate).
In the medium SNR, we see that the rate is close to the cases with one-bit quantization. In the high SNR regime, however, the rate will converge to $\log_2(2^{2\Nt})=2\Nt$ bps/Hz, instead of $2\Nr$ bps/Hz in the cases of one-bit quantization.

In Figs. \ref{fig:Rate_Nr_2_Nt_2_low_SNR}, \ref{fig:Rate_Nr_2_Nt_4_low_SNR} and \ref{fig:Rate_Nr_3_Nt_2_low_SNR}, the achievable rates in the low SNR regime (below $-10$ dB) are plotted versus the SNR in linear scale.
The slopes of these curves verify our analysis.
For independent QPSK signaling, we plot the low SNR capacity approximation given by \eqref{eq:QPSK_1bit_MIMO}.
When $\Nr=\Nt=2$, we find that the channel inversion method is worse than the QPSK signaling. However, the channel inversion method is better when $\Nr = 2, \Nr =4$.
This verifies our analysis in Section \ref{sec:channel_inversion}.

As shown in Figs. \ref{fig:Rate_Nr_2_Nt_2}-\ref{fig:Rate_Nr_3_Nt_2_low_SNR}, optimizing the probabilities by the Blahut-Arimoto algorithm does not improve the achievable rates at the high SNR but provides gain at low and medium SNR. In addition, the convex optimization method with probabilities optimized by the Blahut-Arimoto algorithm achieves the largest rate in the cases with one-bit quantization.



\subsection{MmWave Channel with One-Bit Quantization}
\begin{figure}[t]
\begin{centering}
\includegraphics[scale=.6]{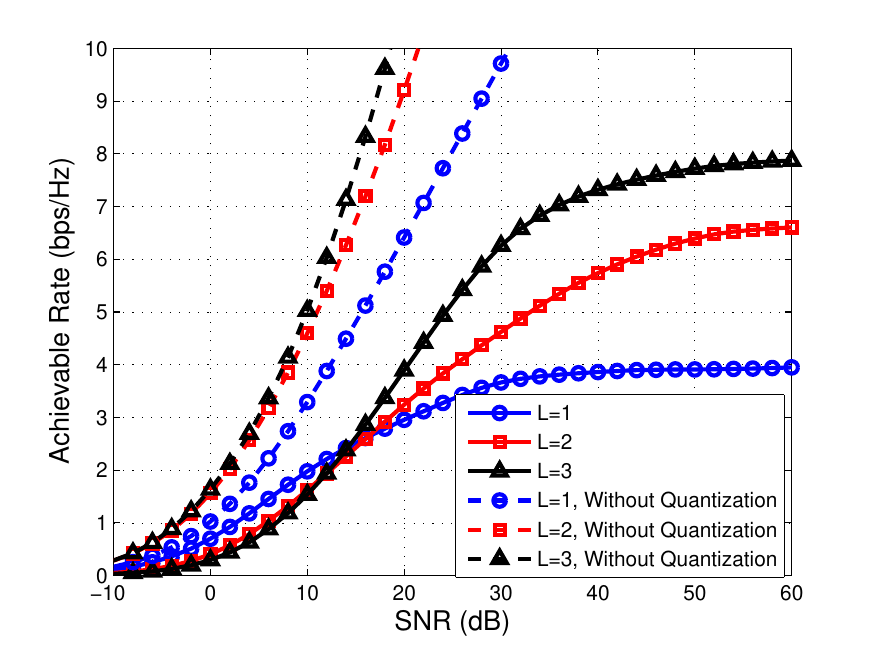}
\vspace{-0.1cm}
\centering
 \caption{The achievable rate of the mmWave $4 \times 256$ mmWave system with different number of paths. $2 \times 2$ and $16 \times 16$ planar antenna arrays are installed at the receiver and transmitter, respectively. The inter-element spacing is one half of the wavelength.}\label{fig:mmWave_Rate}
\end{centering}
\vspace{-0.3cm}
\end{figure}

We show the achievable rates in a $4 \times 256$ channel with varying number of paths in Fig. \ref{fig:mmWave_Rate}.  The azimuth angles $(\varphi_{\mr{t}}, \varphi_{\mr{r}})$ and elevation angles $(\theta_{\mr{t}}, \theta_{\mr{r}})$ are uniformly distributed over $[0, 2\pi]$ and $[-\pi/2, \pi/2]$, respectively. The complex path gains $\alpha$ are complex Gaussian variables. The inter-element spacing of the receiver antenna array is set to one half of the wavelength.
It is shown that as $L$ increases from $1$ to $3$, the achievable rate at high SNR increases. The rates converge to $\log_2 K(4, L)$ bps/Hz, which is $4$, $7$ and $7.9$ bps/Hz when $L=1$, $2$, $3$, respectively. In addition, the rates only converge at very high SNR which is larger than 60 dB in the figure.
The rates of the mmWave channel without quantization are also shown.
We see that the loss incurred by the use of one-bit ADCs increases with the number of paths $L$.

\section{Conclusion}
In this paper, we analyzed the capacity of point-to-point $\Nr \times \Nt$ MIMO channel with one-bit quantization at the receiver. CSI was assumed to be known at both the transmitter and receiver. We obtained the MISO channel capacity in closed-form. For the SIMO and MIMO channel, we derived bounds on the capacities at infinite and finite SNR. A convex optimization based method was also proposed to design the transmitter constellation. Last, we considered the mmWave MIMO channel with limited path and showed the capacity is limited by the number of paths.

From the main results present in this paper, we draw several conclusions.
First, when there is single antenna at the receiver, the capacity is achieved by precoding and QPSK signaling.
Second, the MIMO channel capacity at infinite SNR is related to a classic combinatorial geometric problem.
Our proposed input design method can approach the infinite SNR capacity.
Third, when the channel matrix has full row rank and a small condition number, channel inversion precoding is close to the optimum.
Last, treating the quantization error as Gaussian noise is unsuitable at high SNR.

There are many potential directions of future work. Perhaps the most critical assumption is that the transmitter and receiver have complete and perfect CSI. Channel estimation with one-bit quantization has been considered in \cite{Dabeer_ICC10, Zeitler_TSP12, Lok_ISIT98, Ivrlac_WSA07, Mezghani_WSA10} at lower frequencies.
Due to the sparse structure in mmWave channels, it is of interest to develop efficient compressed channel estimation techniques at the receiver and also  methods for feeding back CSI to the transmitter. Initial work on channel estimation using the one-bit compressive sensing framework has been reported in \cite{Boufounos_CISS08, Mezghani_WSA12, Jacques_IT13, Wen_Chao-Kai_arxiv15} and our work in \cite{Mo_Jianhua_Asilomar14}. Another possible direction is to consider the effects of imperfect CSI. With imperfect CSI, the design of the transmitter constellation should be robust to the CSI error term. Studying the effect of imperfect CSI is left to future work.

\appendices
\section{Bounds of the Achievable Rate of Convex Optimization Method}\label{app_lowerbound}
We provide a lower bound for the achievable rate of the convex optimization strategy. Assume that there are $M$ convex optimization problems having nonzero objective function values. Denote the quantization vectors, optimal solutions and optimal objective function values of these $M$ problems as $\widehat{\mb{r}}^i$, $\widehat{\mb{x}}^i$ and $d^i$, $1 \leq i \leq M$, respectively. Denote the minimum $d^i$ as $d_{\min}$.
Assume that these $M$ symbols are transmitted with equal probability $\frac{1}{M}$. The conditional entropy of $\widehat{\mb{r}}$ is,
\begin{eqnarray*}
  \cH( \widehat{\mb{r}}|\widehat{\mb{x}})&=&  \sum_{i=1}^M \mr{Pr}(\widehat{\mb{x}}^i) \cH(\widehat{\mb{r}}|\widehat{\mb{x}}^i)  \\
  &\stackrel{(a)}=& \frac{1}{M} \sum_{i=1}^M  \sum_{j=1}^{2 \Nr} \cH \left( \widehat{\br}_j | \widehat{\bx}^i \right) \\
  &\stackrel{(b)}=& \frac{1}{M}\sum_{i=1}^M  \sum_{j=1}^{2 \Nr} \Hb \left( Q \left( \sqrt{2} \left(\widehat{\bH} \widehat{\bx}^i \right)_j \widehat{\mb{r}}^i_j  \right) \right) \\
  &\stackrel{(c)}{\leq} & \frac{1}{M} \sum_{i=1}^M \sum_{j=1}^{2 \Nr} \Hb \left( Q \left(\sqrt{2} d^i \right) \right) \\
  &\leq &  \frac{1}{M} \sum_{i=1}^M \sum_{j=1}^{2 \Nr} \Hb \left( Q \left(\sqrt{2} d_{\min} \right) \right) \\
  &=& 2 \Nr \Hb\left( Q \left( \sqrt{2 d_{\min}^2} \right)\right)
\end{eqnarray*}
where $(a)$ follows from that the noises across different antennas are independent, $(b)$ follows from that the in-phase and quadrature parts of the noises are independent and Gaussian distributed with variance $\frac{1}{2}$,  $(c)$ follows from that $\left(\widehat{\bH} \widehat{\bx}^i \right)_j \widehat{\mb{r}}_j \geq d^i$.

For $1\leq i \leq M$, we have
\begin{eqnarray}
  \Pr(\widehat{\mb{r}}^i) &=& \sum_{j=1}^{M} \Pr(\widehat{\mb{x}}^j) \Pr(\widehat{\mb{r}}^i|\widehat{\mb{x}}^j) \\
  &\geq & \Pr(\widehat{\mb{x}}^i) \Pr(\widehat{\mb{r}}^i|\widehat{\mb{x}}^i) \\
  &\geq & \frac{1}{M} \left(1-Q\left(\sqrt{2 d_{\min}^2}\right)\right)^{2\Nr}.
\end{eqnarray}
We denote $q := \left(1-Q\left(\sqrt{2 d_{\min}^2}\right) \right)^{2\Nr}$.
We now consider the following problem,
\begin{subequations}
\begin{eqnarray*}
  \min & &\cH(\widehat{\mb{r}}):=-\sum_{i=1}^{2^{2\Nr}} \Pr(\widehat{\mb{r}}^i) \log_2 \Pr(\widehat{\mb{r}}^i)  \\
   \text{s.t.} &  & \mr{Pr}(\widehat{\mb{r}}^i) \geq \frac{q}{M}, \quad 1\leq i \leq M, \\
   & & \mr{Pr}(\widehat{\mb{r}}^i) \geq 0, \quad M < i \leq 2^{2\Nr}, \\
   & & \sum_{i=1}^{2^{2\Nr}} \mr{Pr}(\widehat{\mb{r}}^i)=1.
\end{eqnarray*}
\end{subequations}
Since the entropy function is concave \cite{Cover_Book12}, the minimum is achieved at the extreme points.
There are two kinds of extreme points.
\begin{enumerate}
  \item The first kind of extreme points:
  \begin{equation}
        \left\{
        \begin{array}{r@{\;=\;}l}
        \mr{Pr}(\widehat{\mb{r}}^j) & 1-\frac{(M-1)q}{M}, \text{for a } j \; \text{satisfying } 1 \leq j\leq M,\\
        \mr{Pr}(\widehat{\mb{r}}^i) & \frac{q}{M}, \quad \text{for all } 1 \leq i \leq M \text{ and } i \neq j, \\
        \mr{Pr}(\widehat{\mb{r}}^i) & 0, \quad \text{for all } M < i \leq 2^{2 \Nr},
        \end{array}
        \right.
        \end{equation}
  \item The second kind of extreme points:
        \begin{equation}
        \left\{
        \begin{array}{r@{\;=\;}l}
        \mr{Pr}(\widehat{\mb{r}}^i) & \frac{q}{M}, \quad \text{for all }  1 \leq i \leq M\\
        \mr{Pr}(\widehat{\mb{r}}^j) & 1-q, \text{for a } j \; \text{satisfying } M \leq j\leq 2^{2 \Nr},\\
        \mr{Pr}(\widehat{\mb{r}}^i) & 0, \quad \text{for all } M < i \leq 2^{2 \Nr} \text{ and }i \neq j
        \end{array}
        \right.
        \end{equation}
\end{enumerate}
The corresponding entropies of $\widehat{\mb{r}}$ are,
\begin{eqnarray*}
  a_1&:=&-(M-1) \frac{q}{M}\log_2 \frac{q}{M} \\
  & & \quad -\left(1- \frac{(M-1)q}{M} \right) \log_2 \left(1- \frac{(M-1)q}{M} \right), \\
  a_2&:=&-q\log_2 \frac{q}{M}-\left(1-q \right) \log_2 \left(1- q \right),
\end{eqnarray*}
respectively.

Therefore, the mutual information is
\begin{eqnarray*}
  & & I(\mb{x}; \mb{r}) \\
  &=& \cH (\mb{r}) - \cH(\mb{r}|\mb{x}) \\
  & \geq & \min\{a_1, a_2\} - 2 \Nr \Hb\left( Q \left( \sqrt{2 d_{\min}^2} \right)\right)
\end{eqnarray*}
In $\textbf{P2}$, we can see that $d_{\min}^2 = \alpha(\bH) \Pt$ where $\alpha(\bH)$ is a constant depending on the channel $\bH$.

Last, an upper bound of $d_{\min}$ can be derived as follows,
\begin{eqnarray}
    2 d_{\min}^2  & \leq & \min_{i,j} \|\left(\mb{H} \mb{x}^i \right)_j \|^2 \leq \Pt \left( \min_{1\leq i \leq \Nr} \|\mb{h}_i\| \right).
\end{eqnarray}
Hence,
\begin{eqnarray}
  d_{\min}^2 \leq \frac{1}{2} \Pt \left( \min_{1\leq i \leq \Nr} \|\mb{h}_i\| \right).
\end{eqnarray}

\bibliographystyle{IEEEtran}
\bibliography{IEEEabrv,One_bit_quantization}

\end{document}